\documentclass[useAMS,usenatbib]{mn2e}
\usepackage{epsfig,graphicx,latexsym,amsmath,amssymb}
\usepackage{natbib}
\usepackage{mathrsfs}
\usepackage{lastpage}
\usepackage{multirow}
\usepackage[hyphens]{url}
\usepackage{hyperref}
\usepackage{color}
\usepackage{colortbl}
\definecolor{kugray5}{RGB}{224,224,224}

\bibpunct[,]{(}{)}{;}{a}{}{,}

\usepackage{color}

\title[Mergers of multi-metallic globular clusters]
       {Mergers of multi-metallic globular clusters:\\
        The role of dynamics}
\author[P. Amaro-Seoane et al] 
{Pau Amaro-Seoane$^{1}$
                        \thanks{E-mail: Pau.Amaro-Seoane@aei.mpg.de (PAS)},
Symeon Konstantinidis$^{2,\,3}$,
Patrick Brem$^{1}$ \& M\'{a}rcio Catelan$^{2}$
   \\
$^{1}$Max Planck Institut f\"ur Gravitationsphysik
(Albert-Einstein-Institut), D-14476 Potsdam, Germany\\
$^{^2}$ Pontificia Universidad Cat\'olica de Chile, Facultad de F\'{i}sica,
        Departamento de Astronom\'\i a y Astrof\'\i sica,\\
         Av. Vicu\~{n}a Mackenna 4860, 782-0436 Macul, Santiago, Chile
\\
$^{^3}$Astronomisches Rechen-Institut, M{\"o}nchhofstra{\ss}e 12-14, 69120,
Zentrum f\"ur Astronomie, Universit\"at Heidelberg, Germany
}

\begin{document}

\date{draft \today}

\pagerange{\pageref{firstpage}--\pageref{lastpage}} \pubyear{2013}

\maketitle

\label{firstpage}

\begin{abstract}
{\em Hubble Space Telescope} observations of globular clusters (GCs) in the
Antenn{\ae} galaxy show clusters of clusters, or regions in the galaxy that
span hundreds of parsecs, where many of the GCs are doomed to collide, and
eventually merge. Several such objects appear likely to present a significant
range in ages, hence possibly metallicities, and their merger could plausibly lead to multi-metallic GCs.
Here we explore this process with direct-summation $N-$body simulations with
GPU hardware.  Our results reveal that colliding GCs with different
metallicities and ages can produce a GC with multiplicity and occupation
fractions not unlike those observed in multi-metallic clusters. In our
simulations, the merged clusters have a phase with a larger amount of
flattening than average, as a consequence of rapid rotation~-- thus suggesting
that relatively recent mergers may play a role in producing highly flattened,
multi-metallic clusters.  We additionally explore the role of the King parameter of the
cluster in the occupation fractions with a set of 160 direct-summation
simulations and find that for equal size clusters the King parameter of the
progenitor clusters determines the occupation fractions in the merger product,
while in unequal size mergers the size of the clusters dominates the
distribution of stars in the new GC.
{In particular, we find that the {\em observed} distribution of populations in $\omega$ Cen can be described
to some extent with our dynamical models.}
\end{abstract}


\section{Introduction}
\label{sec.motivation}

The merger history of globular clusters (GCs) is increasingly being
recognized as an important aspect of GC research
\citep[e.g.,][ and references therein]{svb96,adea00,DMG02,dmea04,LaneEtAl2010,by12,PeacockEtAl2013}.
In the case of the Milky Way,
a merger history has most recently been suggested for NGC~1851, on the basis
of detailed spectroscopic analysis of a large sample of red giants
\citep{CarrettaEtAl2010c,CarrettaEtAl11}. Indeed, the existence of a
small but non-negligible metallicity spread in this cluster had also been
suggested previously by \cite{LeeEtAl2009}, who first hinted that the
cluster's metallicity distribution may actually be bimodal.

As pointed out by \cite{mc97}, the presence of bimodal, or even multi-modal,
metallicity distributions is generally expected, in the case of the GC merger
scenario originally envisaged by \cite{svb96}. On the other hand, the empirical
evidence suggests that mergers of clusters of different metallicities may have
been few in the Milky Way and its immediate vicinity \citep{mc97}. However, the
situation may be more favorable in other environments.  In particular, in the
Antenn{\ae} galaxy (NGC~4038/NGC~4039), gravitationally bound clusters appear
to exist with a sizeable difference in ages, hence possibly also metallicity.
These objects will eventually merge~-- and, if they survive long enough, will
eventually lead to GC-like objects with bimodal, and possibly multi-modal
distributions \citep[e.g.][]{Kroupa98,PeacockEtAl2013}.
In a more general sense, mergers of star clusters may play an important role in
the explanation of the complex abundance patterns that are observed in Galactic
GCs, which include not only metallicity spreads in massive systems like $\omega
$~Centauri (NGC~5139) and M54 (NGC~6715), but also evidence of multiple
populations, as indicated by the abundances of chemical species such as O, Na,
Mg, Al, and also the observed color-magnitude diagrams
\citep[e.g.][and references therein]{CarrettaEtAl10,CarrettaEtAl11,Bekki11,Bekki2012,JooLee2013}.
In this
paper, we will explicitly tackle only the aforementioned {\em global}
metallicity variations.

The purpose of this paper is accordingly to provide
the first numerical simulation of a GC merger involving components of different
ages and metallicities. Indeed, while other simulations of GC mergers have been
carried out by other authors, including \cite{MakinoEtAl91}, \cite{Hurley03}),
\cite{adea00}, \cite{ct01}, \cite{pwr07}, and \cite{dlfm210}, to the best of
our knowledge, ours is the first study to explicitly track the metallicity of
the individual stars in the course of the merger \citep[see
also][]{AmaroSeoaneEtAUCDs}.

Our paper is structured as follows. In \S\ref{sec.ansatz} we provide more details
regarding the properties of potential merger progenitors in the Antenn{\ae} galaxy,
as direct motivation for our numerical experiments. In \S\ref{sec.numerical},
we describe our numerical simulations, along with
the region of parameter space explored in our calculations. In \S\ref{sec.results}
we present our main results, with particular emphasis on the resulting
color-magnitude diagrams (CMDs) and the change
in cluster shape as a function of time. In \S\ref{sec.results2} we show the impact of
the size and King parameter on the merger product. In sections \S\ref{sec.rot} and \S\ref{sec.omecen} we present an analysis of the rotation of the merged clusters and the particular case of $\omega$ Cen. Finally, in \S\ref{sec.summary} we summarize
our results and present some additional discussion.

\section{Smashing clusters with different metallicities: Ansatz and numerical tools}
\label{sec.ansatz}

Observations of colliding galaxies such as the Antenn{\ae} galaxy show bound
systems of young, massive clusters. In this system, {\em Hubble Space Telescope}
(HST) observations exhibit
relatively small regions spanning a few hundreds of pc embracing hundreds or
even thousands of young clusters, i.e.  clusters of clusters or ``cluster
complexes'' \citep[CC from now onwards, see e.g.][ and references
therein]{Kroupa98,WhitmoreEtAl10}. These have been proposed to be the
progenitors of ultra-compact dwarf galaxies (UCDs) or even massive GCs
as a result of the agglomeration of hundreds of their member clusters
\citep{FellhauerKroupa02,FellhauerKroupa05,BruensEtAl11,BruensKroupa2011,AmaroSeoaneEtAUCDs,Bekki2012,PeacockEtAl2013}.

There are different reasons why
two clusters participating in a collision may have different metal contents.
For instance, Fe is produced during supernova explosions (SNe),
which create a very fast moving gas that cannot be retained in the cluster
because of its shallow potential well \citep[unless the cluster
was born at least 10 times more massive than it is today; see][ and references
therein]{Renzini2008,ValcarceCatelan2011}. In a star forming region, this gas
can mix with gas clouds surrounding the parent cluster and, after slowing and
cooling down, create a new, younger cluster~-- the child cluster~-- with a
characteristic stellar Fe abundance higher than the one characterizing the stars
of the parent cluster.  The child and parent clusters can then merge with one
another, giving birth to another cluster with two well-defined stellar populations:
the metal-poor stars of the parent cluster and the metal-rich stars of the child
cluster.

Indeed, recent detailed observations of some of the Antenn{\ae} galaxy's CCs (such
as the ``knot S'' and ``knot B'') \citep{WhitmoreEtAl10} typically show a single
massive cluster older (habitually by a few tens of Myr) than the rest of the
members of the CC located at the centre of a giant molecular cloud.
\cite{WhitmoreEtAl10} suggest that the characteristic location of the older
cluster and its age might be explained in terms of interactions between the
older cluster and the gas cloud.  The difference in age would be an
indicator that the giant stars in the old cluster have released gas because
of SN explosions, thus polluting the surrounding gas.  Therefore, the stars
of the new clusters must have a clear-cut different Fe abundance.
These CCs might accordingly be a natural breeding ground for
multi-metallic GCs, because they have different ab initi$\bar{\rm o}$
metallicities, collide, and merge. Indeed, evidence of already merged clusters
in the Antenn{\ae} gaxaxy has been found by \cite{GreisslEtAl2010} who observed
the spectrum of a massive star cluster of this galaxy and concluded that the cluster
is a superposition of a young and an older one with ages below $3$ Myr
and between $6-18$ Myr respectively \citep[see also][]{PeacockEtAl2013}. The observed age-difference may plausibly be accompanied
by a difference in the metallicity of the two populations.

\section{Numerical simulations: Tools and general setup}
\label{sec.numerical}

\begin{table*}
\begin{center}
\begin{tabular}{cc|c|c|c|c|c|}
\cline{3-7}
& & \multicolumn{5}{|c|}{Parameters}                                                    \\ \cline{3-7}
& Cl\# & W0 & $R$ (pc) & $Z$ & Age (Myr) &  Distances          \\ \cline{1-7}
\multicolumn{1}{|c|}{\multirow{2}{*}{${\cal A}$}} &
\multicolumn{1}{|c|}{1} & 9 & 6 & 0.002 & 100 & $R_{\rm COM}=12$                 \\ \cline{2-6}
\multicolumn{1}{|c|}{}                        &
\multicolumn{1}{|c|}{2} & 5 & 3 & 0.001 & 50 &  $d_{\rm min} = 0.5$               \\ \cline{1-7}
\multicolumn{1}{|c|}{\multirow{2}{*}{${\cal B}$}} &
\multicolumn{1}{|c|}{1} & 12 & 5 & 0.006 & 50 &  $R_{\rm COM}=10.6$               \\ \cline{2-6}
\multicolumn{1}{|c|}{}                        &
\multicolumn{1}{|c|}{2} & 5 & 3 & 0.005 & 100 & $d_{\rm min} = 0.5$               \\ \cline{1-7}
\multicolumn{1}{|c|}{\multirow{2}{*}{${\cal C}$}} &
\multicolumn{1}{|c|}{1} & 12 & 5 & 0.02 & 50 &  $R_{\rm COM}=10.6$                \\ \cline{2-6}
\multicolumn{1}{|c|}{}                        &
\multicolumn{1}{|c|}{2} & 5 & 3 & 0.01 & 100 &  $d_{\rm min} = 0.5$               \\ \cline{1-7}
\multicolumn{1}{|c|}{\multirow{2}{*}{${\cal D}$}} &
\multicolumn{1}{|c|}{1} & 9 & 6 & 0.002 & 100 &  $R_{\rm COM}=52.8$               \\ \cline{2-6}
\multicolumn{1}{|c|}{}                        &
\multicolumn{1}{|c|}{2} & 5 & 3 & 0.001 & 10 &  $d_{\rm min} = 2.0$               \\ \cline{1-7}
\multicolumn{1}{|c|}{\multirow{3}{*}{${\cal E}$}} &
\multicolumn{1}{|c|}{1} & 9 & -- & 0.002 & 100 &  $1+2~{\rm from}~ {\cal A}$        \\ \cline{2-5}
\multicolumn{1}{|c|}{}                    &
\multicolumn{1}{|c|}{2} & 5 & -- & 0.001 & 50 & $R_{\rm COM,\,2}=49$               \\ \cline{2-6}
\multicolumn{1}{|c|}{}                        &
\multicolumn{1}{|c|}{3} & 5 & 5 & 0.009 & $2.1\times 10^3$ & $d_{\rm min,\,2} = 10$ \\ \cline{1-7}
\end{tabular}
\end{center}
\caption{
Initial conditions for the clusters. From the left to the right we have the
simulation ID, the cluster number (Cl\#), the $W_0$ parameter, the initial
radius of the cluster, the metallicity and the initial age of the cluster.
On the last right
column we show the initial distance between the center-of-mass of the two
clusters that collide, $R_{\rm COM}$ (which corresponds to $|{\bf x_1}+{\bf
x_2}|$ of Fig. 1 in Amaro-Seoane 2006), and $d_{\rm min}$, both in pc. In
all simulations but ${\cal E}$ we use an initial total number of stars of
60,000 (30k for each cluster).  In case ${\cal E}$ we use the outcome of ${\cal
A}$, which is a cluster of 52,691 stars, and a radius of 20 pc and make it
collide with a third cluster, of $N_{\star} = 20,\!000$ stars for ${\cal E}$. In
this simulation we add the new distance between COM, $R_{\rm COM,\,2}$ and
$d_{\rm min,\,2}$, for the second collision. We note that the metal-rich cluster
is initially modeled with a higher value of $W_0$ because the observations
suggest that the metal-rich population may be more centrally concentrated.
}
\label{tab.InitData}
\end{table*}

We run direct-summation
$N-$body simulations of clusters with initially different metallicity contents
and ages to investigate this and analyse the impact of dynamics on the
occupation fractions of the different populations in the CMDs.

We set initially the clusters on a parabolic orbit so that the minimum distance
at which they pass by is $d_{\rm min}$ if they are considered to be point
particles at their centers of mass, as described in \cite{Amaro-Seoane06}.
 We use a \cite{Kroupa01} initial mass function
for the stars of the clusters with mass limits 0.2 $\rm{M}_\odot$ and
100 $\rm{M}_\odot$.
 We integrate the evolution with direct-summation $N-$body tools, which integrate
all gravitational forces for all particles at every time step, without making
any a priori assumptions about the system. The code we have employed, {\sc
NBODY6-gpu}, uses the improved Hermite integration scheme
\citep{Aarseth99,Aarseth03}. This needs computation of not only the
accelerations, but also their time derivatives. The programme also includes
{\em KS regularisation} and {\em chain regularisation}, so that when particles
are tightly bound or their separation becomes too small during a hyperbolic
encounter, the system is regularised \citep{KS65,Aarseth03} to prevent too
small individual time steps.

We ran different cases with different initial conditions, as shown in
Table~\ref{tab.InitData}\footnote{See also \url{http://members.aei.mpg.de/amaro-seoane/ASKBC}}.
The clusters, assumed to be isolated, were modelled initially with a King model
of different concentrations W0 \citep{ik66}, radii and metallicities and
evolved for different times with the stellar evolution package {\tt sse},
described in \cite{HurleyEtAl00}.  For the particular case ${\cal E}$, we use
the outcome of simulation ${\cal A}$ and make it merge with another cluster
(numbered 3) on a parabolic orbit with a new $d_{\rm min,\,2}$ and $R_{\rm
COM,\,2}$, as indicated on the right of Table~\ref{tab.InitData}.  During the
collision we neglect stellar evolution, because in all runs it took
approximately a few Myr and the impact of evolving the masses on the global
dynamics is negligible. We find in our simulations a significant mass loss
after the merger of the clusters that affects the different occupational
fraction numbers of the CMDs.

\section{Results: Different dynamical and chemical configurations}
\label{sec.results}

\subsection{CMDs and fractional occupation numbers}
\label{sec.CMD}

To determine when the clusters merge, we locate the density centers of the two
clusters and that of the merged system following \cite{CasertanoHut85}.
We follow the simulations after the density centres have coincided for about
one half-mass relaxation time $T_{\rm rlx,\,h}$ of the final cluster.
This allows us to study the distribution of stars due to the dynamics
of the system, which is important to understand the impact of mass loss in the CMD
and occupational fractions in different shells of mass around the density
center of the merged system. Naturally, the CMD we obtain from the simulations
corresponds to idealised observational conditions. In real observations the
measurements are affected by both random and systematic errors and completeness.
Photometric errors are not constant with apparent magnitude. In
general, more luminous stars have smaller photometric errors. This
is described in Fig. 5 of \cite{StentsonEtAl2005} and also in Table 2
of \cite{PerinaEtAl2009}, which show the exponential growth of the photometric
errors with apparent magnitude. Also, faint stars are hard to detect, so
the CMD suffers from completeness (i.e. only a fraction of the faint stars
are actually observed). A description of the completeness in the CMD of real
clusters is given in Fig. 1 of \cite{BrownEtAl2003} and also in
 Fig. 9 and Table 7 of \cite{BuonannoEtAl1994}.
For the transformation of our theoretical CMD to the ``real'' CMD that one
would observe, we introduce an exponential function to guide the
photometric errors, which are then selected from a Gaussian random distribution. We also
included a function for the completeness of the CMD, assuming completeness 100\%
for stars with high luminosities and 0\% for low-luminosity stars.

\begin{figure*}
\resizebox{\hsize}{!}
           {\includegraphics[scale=1,clip]{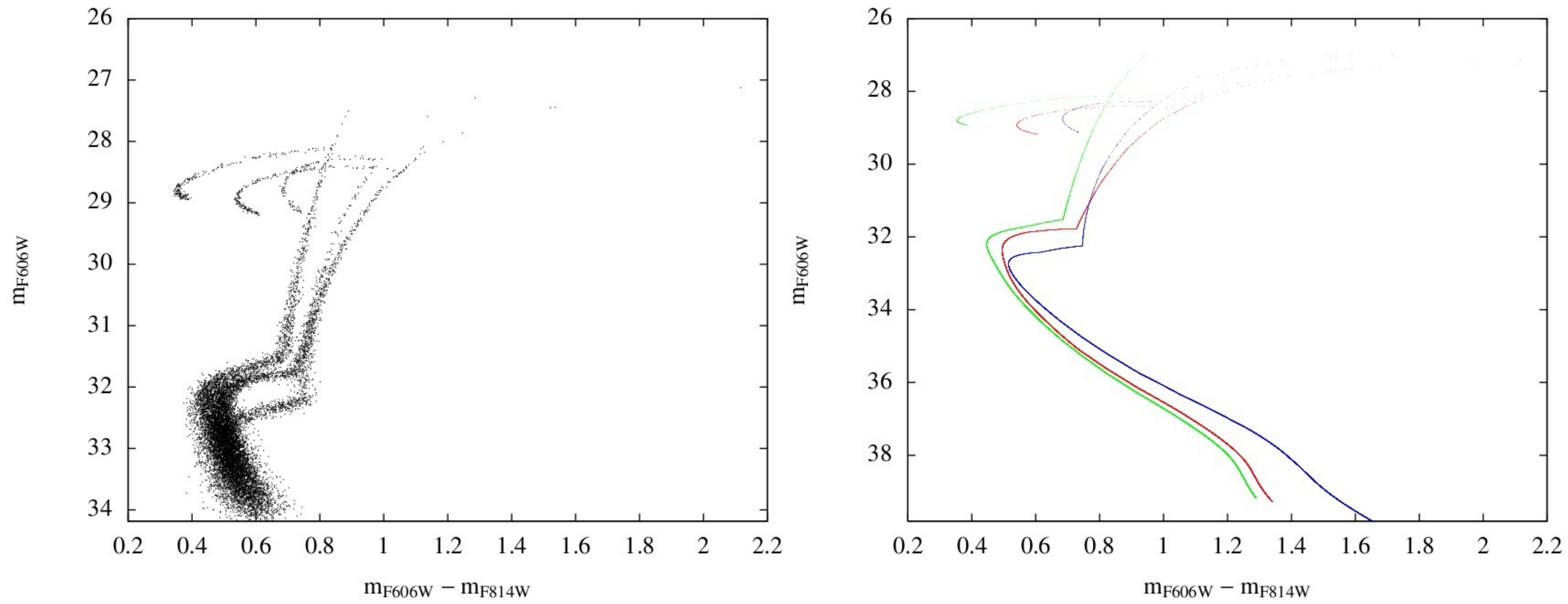}}
\caption
   {
CMD of case ${\cal E}$, a collision of a cluster with another cluster which is
itself the result of a merger of two clusters (case ${\cal A}$), with artificial
errors (left panel) and without them (right panel). The first MS (curve with
the TO point at $\sim 0.45$~mag in $m_{\rm F606W} - m_{\rm F814W}$ in the right
panel, green line in the on-line version of the article) corresponds to the
stellar population of $Z=0.002$, the next MS to $Z=0.001$ (red curve of right
panel) and the third one to $Z=0.009$ (blue curve). The clusters that
harboured the two first metallicities had an age of 100 and 50~Myr
respectively, and the third one had an age of $10^3$~Myr.  The CMD is for
stars in the shell $0 \leqslant r/{\rm pc} < 2.5$ after evolving the merged
cluster for 10~Gyr.
We set the distance of $5$~Mpc
to convert to apparent magnitudes.
The left panel shows the observable CMD of the same cluster.
   }
\label{fig.CMD_E}
\end{figure*}

In Figure~\ref{fig.CMD_E} we show the theoretical and observational CMD of simulation ${\cal E}$,
assuming that the clusters are at a distance of 5 Mpc. As it is obvious, just a fraction of the
upper part of the CMD of such a cluster could be observed with future telescopes,
 but the signature of the merger of three clusters would still be detectable.
 We can see that at the level of the turn-off point (TO) the SGB splits into
three well-defined branches, resembling the CMDs presented by \cite{mc97}.
While such CMDs are not typically found in the Milky Way and its immediate vicinity, they will likely be more commonplace in the
Antennae galaxy's GC system, several Gyr from now. This, in turn, suggests that other galaxies with more violent past histories than
the Milky Way may also be more likely to harbor such GC systems. Candidate hosts for such multi-metallic merged clusters, in this
sense, may include M82  \citep[e.g.][]{KetoEtAl2005,WeiEtAl2012}, at a distance of $3.4\pm 0.1$~Mpc \citep{DalcantonEtAl2009},
and even M31 itself \citep[e.g.][and references therein]{Brown2006,Brown2009,HammerEtAl2010},
at a distance of $752\pm 27$~kpc \citep{RiessEtAl2012}.

We then calculate the occupation ratio of the different stellar
populations, $N_1/N_2$ (and $N_3/[N_1+N_2]$, if we had three different
populations) for different shells starting at the density center. The results
are shown in Table~\ref{tab.OccupFrac}. The
distribution of the different populations in the final, merged cluster
depends mostly on the initial size, concentration, metallicities, and also on the
initial number of stars in each cluster, as well as their initial mass functions
and ages. In most of our models (all but ${\cal B}$ and ${\cal C}$), the metal-rich
cluster, which is always cluster 1, appears to contribute fewer stars to the center
of the merged cluster than the metal-poor population. This appears to be a natural
result of the initial parameters chosen and more specifically of the
different initial size of the clusters, as we will see below.

$\omega$~Cen is the system for which we have the best observational data
about the radial distribution of its multiple populations. \cite{BelliniEtAl09}
do a detailed study showing that stars belonging to the blue MS (bMS) appear
to be more centrally concentrated than stars of the red MS (rMS), with the
fraction $N_{\rm bMS}/N_{\rm rMS} < 1$ outside the core of the system. Since
the bMS contains stars with greater metallicity, according to the authors,
stars with greater metallicity appear to be more centrally concentrated. As
shown in Table~\ref{tab.OccupFrac}, one of our models (${\cal B}$) shows a
distribution in which the metal-rich (MR)
stars appear more centrally concentrated relative
to the metal-poor (MP) stars of the merged cluster. In the same model, MP stars
appear to dominate the region 1-3 pc, while metal-rich stars again dominate
the external shells of the cluster as a result of the large difference in the
initial concentration of the two clusters and in their initial size.
We caution the reader, however,
that $\omega$~Cen may be a much more complex case than described by our models,
since its progenitor system must have been much more massive in the past
\citep[e.g.,][ and references therein]{Renzini2008,ValcarceCatelan2011}, and
the present-day $\omega$~Cen is characterized by a broad, continuum metallicity
distribution, which we are not in a position to properly describe with our
relatively simple $N$-body models, which imply sharply-peaked (if multimodal)
metallicity distributions.

In this context, our models may fare somewhat better in the case of NGC~1851.
In this case, \cite{CarrettaEtAl10,CarrettaEtAl11} observed evidence of a difference
in the distribution of the two populations of red giants. According to them,
the MP component seems to be more centrally concentrated than the
MR one. This observational evidence, which however remains somewhat
controversial \citep{MiloneEtAl09b,CarrettaEtAl11}, contradicts a scenario
for the formation of the second population of stars within the same cluster
\citep[e.g.,][]{Bekki11,ValcarceCatelan2011}, which foresees that the metal-rich
population should be more concentrated around the center. In this sense, the
possibility of a merger origin would appear like an interesting alternative.

\begin{table*} 
\begin{center} 
\begin{tabular}{|c|ccccc|}
\cline{1-6}
 & \multicolumn{5}{c|}{${N}_1/{N}_2$ (and $N_3/(N_1+N_2)$)} \\ \cline{2-6}
Shell (pc) & Case ${\cal A}$ & Case ${\cal B}$ & Case ${\cal C}$ & Case ${\cal D}$ & Case ${\cal E}$ \\
\hline
$0   \leqslant r < 0.5$ &  0.94 & 1.25 & 1.08 & 0.50 & 0.85 (0.32) \\ \cline{2-6}
$0.5 \leqslant r < 1  $ &  0.86 & 1.02 & 1.03 & 0.53 & 0.80 (0.26) \\ \cline{2-6}
$1   \leqslant r < 2  $ &  0.72 & 0.83 & 0.87 & 0.61 & 0.78 (0.23) \\ \cline{2-6}
$2   \leqslant r < 3  $ &  0.66 & 0.81 & 0.82 & 0.90 & 0.74 (0.23) \\ \cline{2-6}
$3   \leqslant r < 4  $ &  0.82 & 1.00 & 0.98 & 1.07 & 0.79 (0.24) \\ \cline{2-6}
$4   \leqslant r < 5  $ &  1.03 & 1.06 & 1.00 & 1.25 & 0.77 (0.27) \\ \cline{2-6}
$5   \leqslant r < 10 $ &  1.39 & 1.19 & 1.17 & 1.52 & 0.95 (0.30) \\ \cline{2-6}
$10  \leqslant r < 50 $ &  1.96 & 1.42 & 1.36 & 2.28 & 1.31 (0.45) \\ \cline{1-6}
\hline
$\epsilon_{0.5,\,\rm max}$ & 0.178 & 0.215 & 0.100 & 0.174 & 0.298 \\ \cline{1-6}
\end{tabular}
\caption{
Occupation fraction and maximum ellipticity
for the mass fraction 0.5, $\epsilon_{0.5,\,\rm max}$ (see caption of Fig.{\ref{fig.Ellipticity_D}}), for the cases of table
\ref{tab.InitData}. The fractions are given in terms of numbers of stars
belonging initially to the first cluster ($N_1$) and the second one ($N_2$) for
different shells of the resulting merged cluster starting from the density
center. For the last two cases we also give the fraction relative to the third
one, $N_3$.
}
\label{tab.OccupFrac}
\end{center}
\end{table*}

\subsection{Ellipticity}
\label{sec.ellipticity}

In study of the parameter space, we have found that the result
of a collision is a cluster that will exhibit phases in the evolution with
$\epsilon$ above average.  In the particular case of simulation ${\cal D}$, the
GC achieves average values after {almost one $T_{\rm rlx,\,h}$},
so that any oblateness would not
be present today in clusters older than their half-mass relaxation time,
unless the collision happened recently
(though we note that we cannot model realistic GCs with our number of stars).
This means that GCs, in particular young ones, with $\epsilon$ above average,
are more likely to harbour populations of stars displaying multi-metallicity;
i.e., any amount of rotation in GCs with multiple metallicities could be a
fingerprint for a dynamical origin.

After the collision, and for a significant fraction of the relaxation time, the
resulting cluster has a significant amount of rotation. This depends on the
initial conditions such as the impact parameter, the \cite{ik66} parameter W0 and the
relative velocity.

\begin{figure}
\resizebox{\hsize}{!}
          {\includegraphics[scale=1,clip]{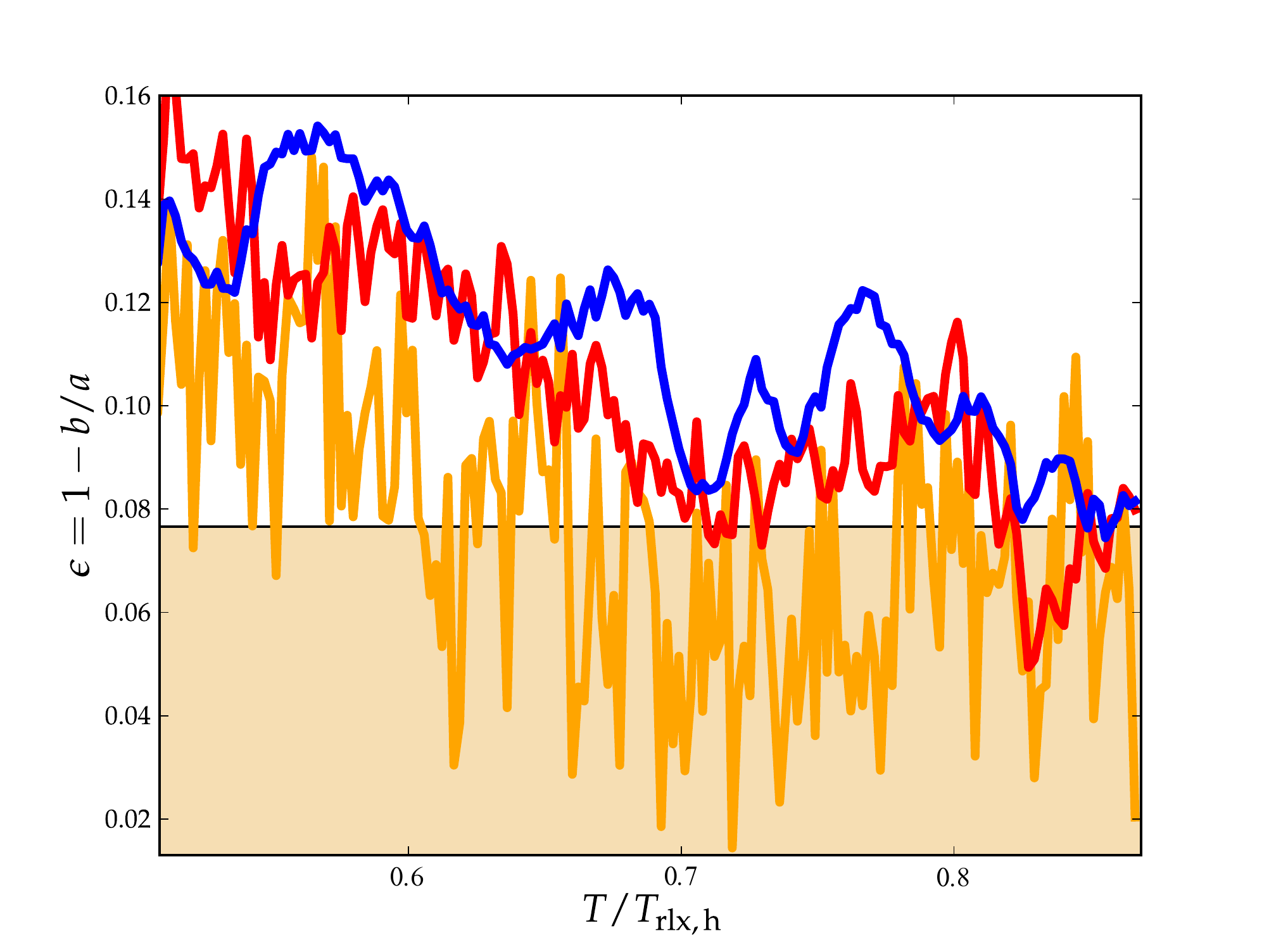}}
\caption
   {
Evolution of $\epsilon$ for case ${\cal D}$ after the density centers coincide.
The semi-major axes are calculated with the ellipsoids of inertia \citep[see
e.g.][]{Chandrasekhar69} and determined by different mass fractions of the
stars (at the lowest value of $T/T_{\rm rlx,\,h}$ in the figure, from the bottom to the top 0.2, 0.5 and 0.8,
displayed in orange, red and blue, respectively in the on-line article).  The
solid black line shows $\hat{\epsilon}$; thus, values within the colored box are
average.  The stars are distributed according to the amount of gravitational
energy; hence, the lower the mass fraction is, the closer we are to the centre
of the resulted merged system.
   }
\label{fig.Ellipticity_D}
\end{figure}

In Figure~\ref{fig.Ellipticity_D} we show $\epsilon$ vs time for ${\cal D}$. We
start the analysis after the three density centers coincide. The system has an
$\epsilon$ above average during a relatively long time and only after
$T \sim 0.9\, T_{\rm rlx,h}$ $\epsilon$
does it reach the average $\sim 0.08$. This suggests that cluster mergers may
lead, at least for almost one $T_{\rm rlx,h}$, to peculiarly flattened systems, whose flattening
may be ascribed to their acquired angular momentum during the merger event.
Consequently, a correlation between multi-metallic stellar populations,
ellipticity and rotation may be expected, in the case of a merger origin.
Note that increased ellipticity was also found in the merger simulations
presented by \cite{doea00} and \cite{ct01}, among others.

Do multi-metallic GCs show systematically high ellipticies and rapid rotation?
Unfortunately, a conclusive answer to this question cannot be provided at present,
given the exceedingly small number of known multi-metallic GCs in the local
Universe. Still, some of the available evidence appears quite suggestive.
$\omega$~Cen, in particular, is well known to be one of the most oblate
Galactic GCs, as also confirmed by the recent, homogeneous measurements
of 116 GCs presented by \cite{ChenChen2010}, based on 2MASS images. $\omega$~Cen
has rotational velocity $v_{\rm rot}/\sigma = 0.32-0.41$, which makes it
one of the fastest rotating GCs in the Galaxy. This could be a signature
for an agglomeration process of a cluster in a CC which receives more and
more impacts from other lighter clusters and ``runs away'' in mass, on its
way to forming a UCD. Measurements of the rotation of the $650$ stars of
$\omega$~Cen \citep{PancinoEtAl07}
show that all subpopulations rotate as a single one
\citep[see also][ for a study of the proper motions of the subpopulations
that supports this result]{AndersonvanderMarel2010}.
This can also be explained in terms
of a collision between GCs, which would assign all stars the same amount of
rotation regardless of their population.

Other Galactic GCs for which a metallicity spread has recently been claimed include M22~=~NGC~6656 \citep[e.g.][]{MarinoEtAl2011,Alves-BritoEtAl2012}
and NGC~2419 \citep{CohenEtAl2010}, both of which are also significantly flattened \citep{ChenChen2010}.
In fact, \cite{BruensKroupa2011}
have suggested that the latter cluster may be the result of a merged star cluster complex,
during the interaction between a gas-rich galaxy and the Milky Way. Very recently, \cite{Bekki2012}
has also considered the possibility
that such clusters may originate from mergers, with the possible production of multi-modal metallicity distributions.

Interestingly, \cite{mgea11} have suggested that the observed
elongation of the starburst cluster Westerlund~1 may similarly be
ascribed to mergers. On the other hand, one should be careful to note
that GC ellipticities may be due to a variety of physical mechanisms, which
are not always easy to disentangle. For instance, \cite{asea06} note that the
GC WLM-1 in the WLM dwarf galaxy in the Local Group, in spite of being one of
the most elongated GCs known, and of being subject to very minor tidal stresses,
does not show any evidence of rotation. We are clearly in face of a very complex
phenomenon, which cannot be explained in terms of any simple scenario. Still,
it does appear like mergers may play an important role in at least some cases,
and we accordingly suggest that the connection between multimodal metallicity
distributions, ellipticity, and rotation be further explored, whenever GC
candidates with multiple metallicities may be detected.

\section{A detailed study of the fractional
occupation number as a function of the King parameter, based on case ${\cal A}$}
\label{sec.results2}

\subsection{Initial data setup}

In order to understand the impact of the King parameter W0 in the final
distribution of different populations as a function of the distance from the
density centre of the merged system, we run a set of 128 simulations with a
setup that has the same initial numerical setup as case ${\cal A}$ of table
\ref{tab.InitData} and the same ages. We explore different parameters, and only
fix the number of stars, the radii, and metallicities, as summarised in Table
\ref{tab.table1}. In this first exploration we set Cluster 2 to have  half the
size of Cluster 1.

\begin{table}
\begin{center}
\begin{tabular}{|c|c|c|c|c|}
\hline
Cluster  & $N$ & $R$ (pc) & $Z$    & Age (Myr)                \\ \hline
         1  & 30001 &   6          & 0.01 &  50                \\ \hline
         2  & 30001 &   3          & 0.04 &  100                 \\ \hline
\end{tabular}
\end{center}
\caption{
Fixed parameters for the detailed analysis of case ${\cal A}$ of the role of the King
paramter. The total mass of the system is 60k $M_\odot$.
}
\label{tab.table1}
\end{table}

We hence vary in the initial data the $W0$ King parameter (i.e. $W0_1$ for
Cluster 1 and $W0_2$ for Cluster 2) and choose the values of 3, 6, 9 and 12.
This gives us 16 possible combinations for the two parameters. For each
combination we run 8 different realizations with initial random seeds to improve
the statistics.
Therefore, the whole set comprehends 128 models.

Initially we set up the clusters as explained in section \ref{sec.numerical}
and we choose a Kroupa initial mass function (IMF) with lower mass $0.2
M_{\odot}$ and higher mass $50 M_{\odot}$ \citep{Kroupa01}.

To evolve the clusters to the assumed initial age, as in the previous section,
we use \textsc{sse}, which uses the metallicity $Z$ and age of each
cluster. Hence, our complete setup for one simulation are two clusters in a
parabolic orbit with their stars evolved to the specified initial age.

\begin{figure*}
\resizebox{\hsize}{!}
          {\includegraphics[scale=1,clip]{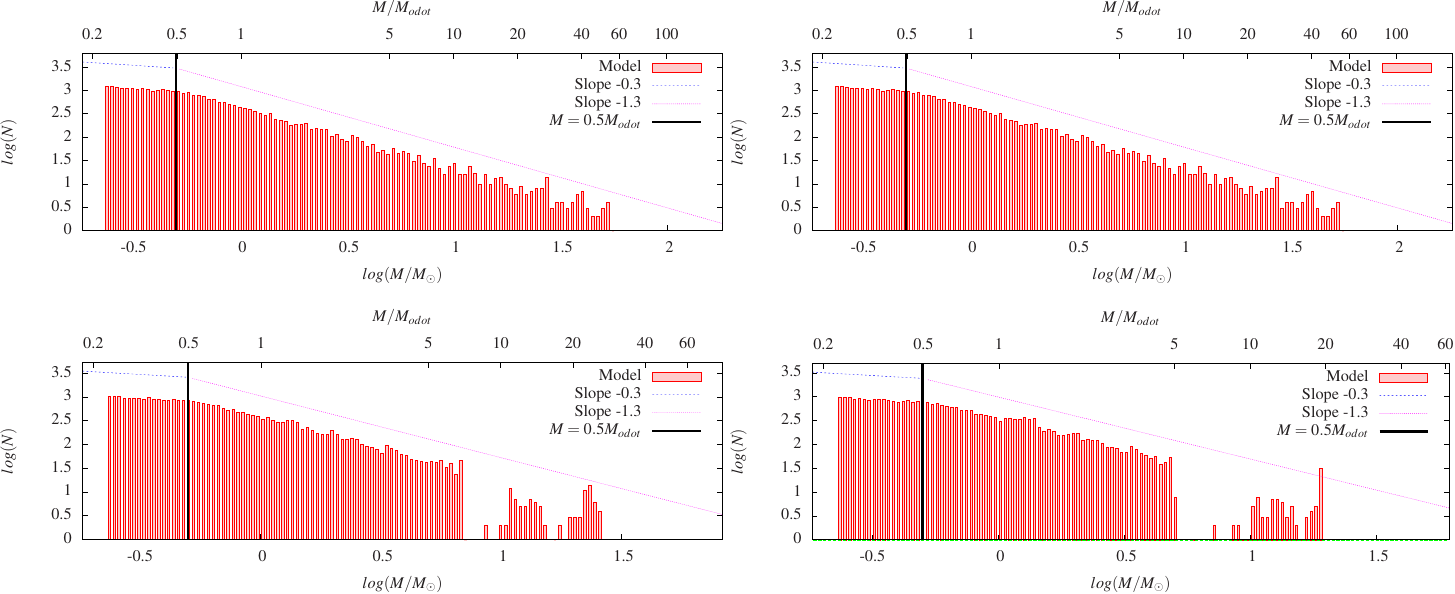}}
\caption
   {
     {\em Upper panels:} IMF of Cluster 1 (left) and Cluster 2 (right) at t=0. The slopes
of the Kroupa IMF are also shown. The transition from one slope
to another happens at 0.5 $M_\odot$.
    {\em Lower panels:} Same at $t=50$ Myr ($Z=0.01$) and $t=100$ Myr ($Z=0.04$)
for cluster 1 and 2, respectively.
   }
\label{fig.IMFs}
\end{figure*}

We can see in figure \ref{fig.IMFs} the mass function of the
two clusters at t=0 and after 50 and 100 Myrs of stellar evolution, as well as the
corresponding slopes.

The lower panels of Fig. \ref{fig.IMFs} show the mass function of the two
clusters after evolving their stars to their initial age. As expected, because
of the small time-scale for stellar evolution ($50$ Myr and $100$ Myr for
Cluster 1 and Cluster 2 respectively), only massive stars have changed
significantly their mass with some of them evolving to BHs with lower mass than
their progenitors.

We run the dynamical evolution of the clusters for
at least one
half-mass relaxation time of the final cluster in all simulations,
typically to $\sim 1.5 \,T_{\rm
rlx,\,h}$.

In table \ref{tab.table_all} we present the results for the whole set of
simulations. We can see that the dependence on the choice for the King
parameter is very weak.  What dominates the evolution in this case is the
difference of sizes. This is why we average over all King parameters in the
table.

We notice that the fraction $N_1$/$N_2$ is smaller than 1 until $r \sim 4$pc in
all simulations.  Below $r = 3$ pc, $N_1$/$N_2$ is smaller than 0.5, meaning
that below this radius the number of stars of Cluster 2 is more than twice
those of Cluster 1.  In the shell $3 < r \leq 5$ the two clusters have almost
equal number of stars.  Finally, in the outer shells of the final merged
cluster, it is stars originally from Cluster 1 which dominate. For $10 < r \leq
50$ Cluster 1 has more than 3 times the stars of Cluster 1.  Since we have
chosen Cluster 1 with twice the size of Cluster 2, the centre of the systems
after merger is more populated with stars that originally belonged to Cluster
2, which was more compact. Accordingly, in the outskirts of the merged cluster
we find that stars from Cluster 1 dominate the population. In our idealised
modelling the systems are isolated but if we added an external galactic
potential, tidal forces would remove more stars of Cluster 1 from the system.

\begin{table*}
\centering
\begin{tabular*}{10cm}{|p{2cm}|p{2cm}|p{2cm}|p{2cm}|p{2cm}|}
\cline{1-5}
Shell  (pc)         & \multicolumn{4}{c|}{$N_1/N_2$}               \\ \cline{1-5}
$0<r\leq 0.5$  &   0.384 $\pm$ 0.083
                        &
                        \multicolumn{1}{p{2cm}|}{\multirow{2}{*}{0.395
                          $\pm$ 0.073}}
                        &
                        \multicolumn{1}{p{2cm}|}{\multirow{4}{*}{0.534
                          $\pm$ 0.074}}
                        &
                        \multicolumn{1}{p{2cm}|}{\multirow{6}{*}{0.639
                          $\pm$ 0.072}}  \\ \cline{1-2}
$0.5<r\leq 1$  &   0.399 $\pm$ 0.071    &   &   &     \\\cline{1-3}

$1<r\leq 2$     &   0.514 $\pm$ 0.081
                        &
                        \multicolumn{1}{p{2cm}|}{\multirow{2}{*}{0.609
                          $\pm$ 0.091}}     & &        \\\cline{1-2}

$2<r\leq 3$     &   0.804 $\pm$ 0.128   & & &           \\\cline{1-4}

$3<r\leq 4$     &   1.150 $\pm$ 0.148
                        &
                        \multicolumn{2}{c|}{\multirow{2}{*}{1.259
                            $\pm$ 0.138}}
                           &        \\ \cline{1-2}

$4<r\leq 5$     &   1.456 $\pm$ 0.156   &  \multicolumn{2}{c|}{\multirow{2}{*}{}}   &            \\\cline{1-5}

$5<r\leq 10$   &   \multicolumn{4}{c|}{\multirow{1}{*}{2.029 $\pm$ 0.349}}                \\\cline{1-5}

$10<r\leq 50$  &  \multicolumn{4}{c|}{\multirow{1}{*}{3.432 $\pm$ 1.164}}                      \\\cline{1-5}

\end{tabular*}
\caption{
Fractional population number for the initial set of simulations in which one of the clusters
has a radius twice as large as the other one, for all King parameters.
}
\label{tab.table_all}
\end{table*}

\begin{figure}
\resizebox{\hsize}{!}
          {\includegraphics[scale=1,clip]{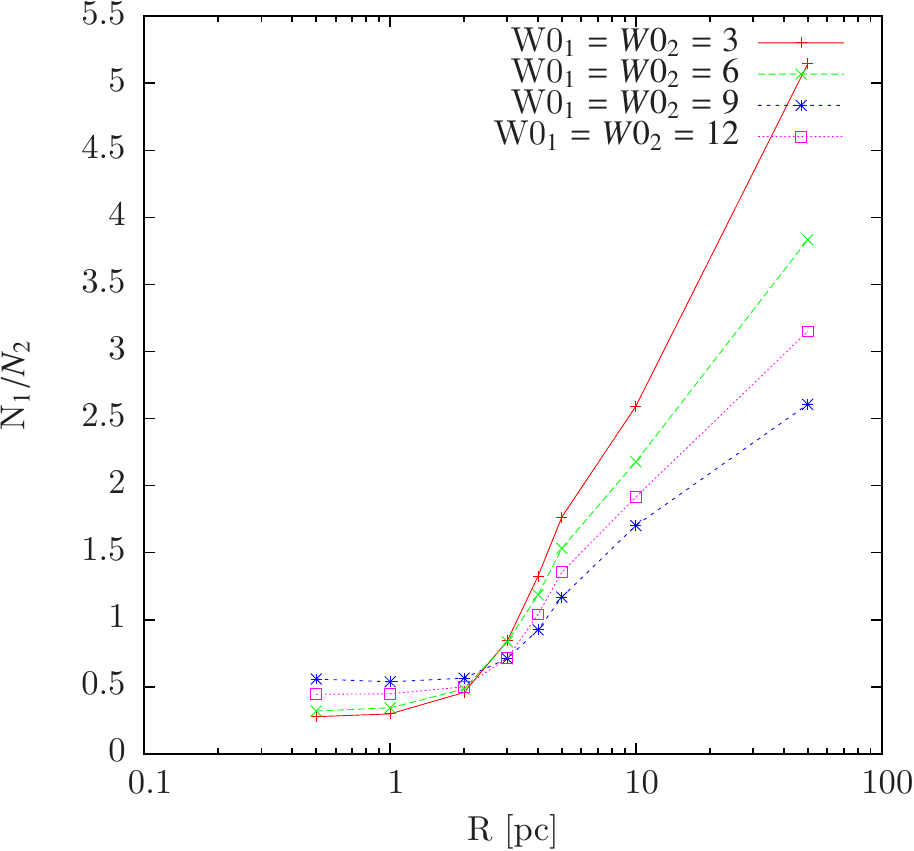}}
\caption
   {
    Fraction $N_1/N_2$ in different shells inside the final cluster at
    T = 1000, for the cases where the two clusters have equal W0 King
parameters and different sizes.
   }
\label{fig.fig_equal_W}
\end{figure}

\begin{figure*}
\resizebox{\hsize}{!}
          {\includegraphics[scale=1,clip]{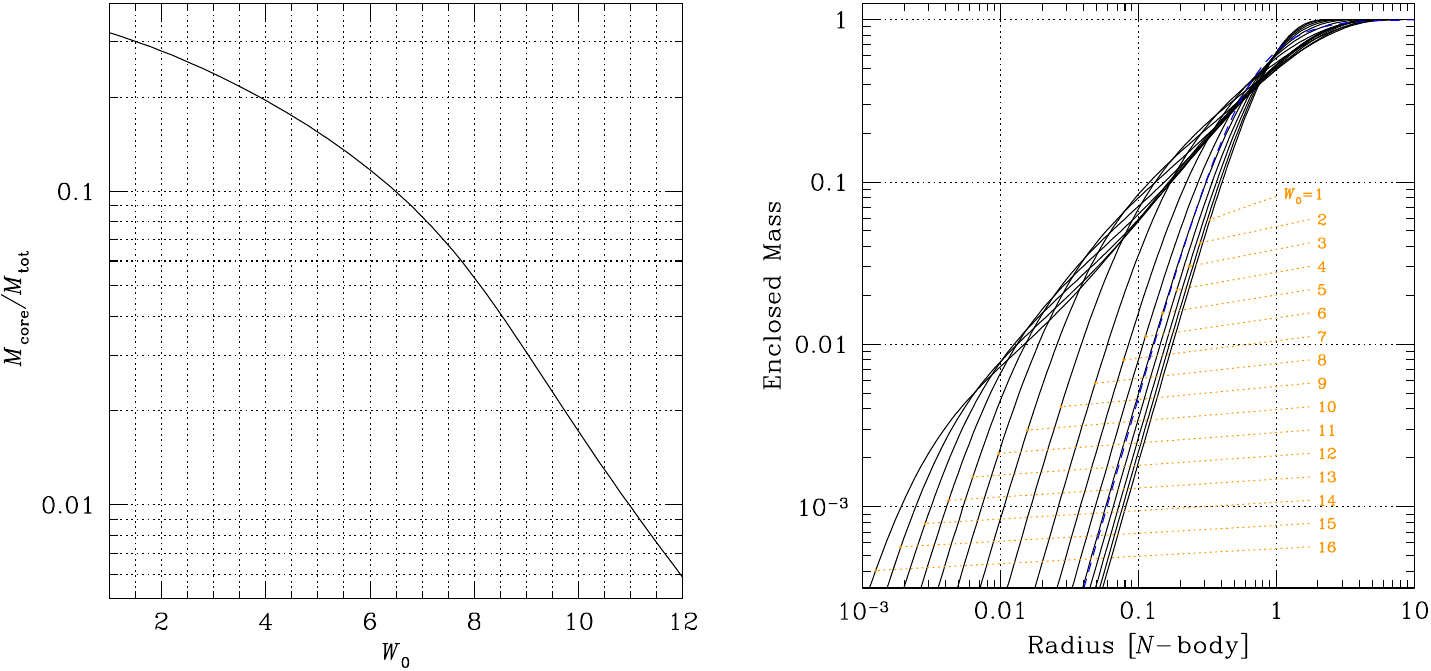}}
\caption
   {
{\em Left panel:} Core mass in units of the total mass as a function of the
King parameter W0. The value W0=12 has a very small core with a tiny fraction
of the total mass in it. {\em Right panel:} Enclosed mass within a certain
radius as a function of that radius for different W0, ranging between 1 and 16.
   }
\label{fig.KingModels}
\end{figure*}

Since the difference in initial sizes plays an important role in the
distribution of stars, in order to understand the impact of the choice for the
initial King parameters we must address the results by first comparing those
simulations which had the same $W0$ for the two clusters.
We hence filter the results of table \ref{tab.table_all} in
which $W0_1=W0_2=3,\,6,\,9,\,12$, respectively, in table
\ref{tab.table_equal_W}. In
Fig. \ref{fig.fig_equal_W} we present graphically the results of this table.
We can conclude from the results that inside a radius $R \sim 3$ pc (which is
close to the half-mass radius), higher W0 parameters lead to higher $N_1/N_2$.
This is true for all cases except W0=12, which has a lower fraction than W0=9,
but still higher than W0=6.  On the other hand, for $R > 3$ pc, an initially
higher W0 parameter leads to a lower fraction $N_1/N_2$, again with the
exception of W0=12. The reason for this is that King models of W0$>$9 have a
very dilute core with very few stars in it. In figure~\ref{fig.KingModels} we
depict the core mass normalised to the total mass and the enclosed mass as a
function of the radius and King parameter.

\begin{table*}
\centering
\begin{tabular*}{14cm}{|p{2cm}|p{3cm}|p{3cm}|p{3cm}|p{3cm}|}
\cline{1-5}
\multicolumn{1}{|p{2cm}|}{\multirow{2}{*}{Shell  (pc)} }       &
  \multicolumn{4}{c|}{$N_1/N_2$}               \\ \cline{2-5}
 & $W0_1 = W0_2 = 3$ &$W0_1 = W0_2 = 6$ & $W0_1 = W0_2 = 9$& $W0_1 = W0_2 = 12$\\ \cline{1-5}
$0<r\leq 0.5$  &   0.276 $\pm$ 0.007
                        &   0.319 $\pm$ 0.016
                        &   0.557 $\pm$ 0.002
                        &   0.443 $\pm$ 0.011\\ \cline{1-5}

$0.5<r\leq 1$  &   0.297 $\pm$ 0.004
                        &  0.342 $\pm$ 0.009
                        &  0.536 $\pm$ 0.006
                        &  0.447 $\pm$ 0.014   \\\cline{1-5}

$1<r\leq 2$     &   0.457 $\pm$ 0.018
                        &   0.482 $\pm$ 0.007
                        &   0.563 $\pm$ 0.019
                        &   0.499 $\pm$ 0.008      \\\cline{1-5}

$2<r\leq 3$     &   0.842 $\pm$ 0.025
                        &   0.834 $\pm$ 0.021
                        &   0.710 $\pm$ 0.014
                        &   0.712 $\pm$ 0.018        \\\cline{1-5}

$3<r\leq 4$     &   1.325 $\pm$ 0.039
                        &   1.185 $\pm$ 0.024
                        &   0.923 $\pm$ 0.029
                        &   1.039 $\pm$ 0.040          \\ \cline{1-5}

$4<r\leq 5$     &   1.765 $\pm$ 0.085
                        &   1.532 $\pm$ 0.052
                        &   1.168 $\pm$ 0.035
                        & 1.353  $\pm$ 0.067\\\cline{1-5}

$5<r\leq 10$   &  2.592 $\pm$ 0.080
                        &  2.177 $\pm$ 0.021
                        &  1.702 $\pm$ 0.070
                        &  1.914 $\pm$ 0.042  \\\cline{1-5}

$10<r\leq 50$  & 5.145 $\pm$ 0.215
                         & 3.833 $\pm$ 0.075
                         & 2.605 $\pm$ 0.083
                         &3.149 $\pm$ 0.119 \\\cline{1-5}

\end{tabular*}
\caption{
Occupation fractional number for the models of table \ref{tab.table_all}
in which initially $W0_1 = W0_2$. A graphical representation of these
results is in figure \ref{fig.fig_equal_W}.
}
\label{tab.table_equal_W}
\end{table*}

\begin{figure*}
\resizebox{\hsize}{!}
          {\includegraphics[scale=1,clip]{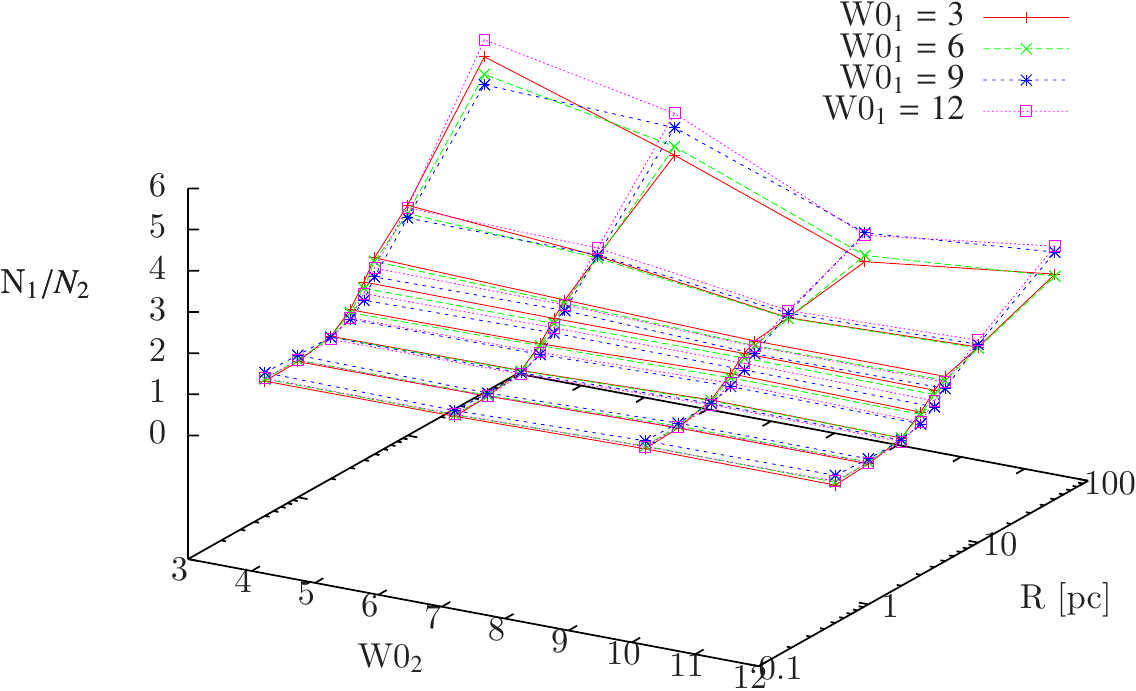}}
\caption
   {
    Fraction $N_1/N_2$ in different shells inside the final cluster at
    about $\sim 1.5 \,T_{\rm rlx,\,h}$,
    for the cases in which $W0_1$ is fixed to 3, 6, 9 and 12,
    while $W0_2$ is free. In these simulations Cluster 1 had twice the size
    of Cluster 2.
   }
\label{fig.fig_W1_constant}
\end{figure*}

\begin{figure*}
\resizebox{\hsize}{!}
          {\includegraphics[scale=1,clip]{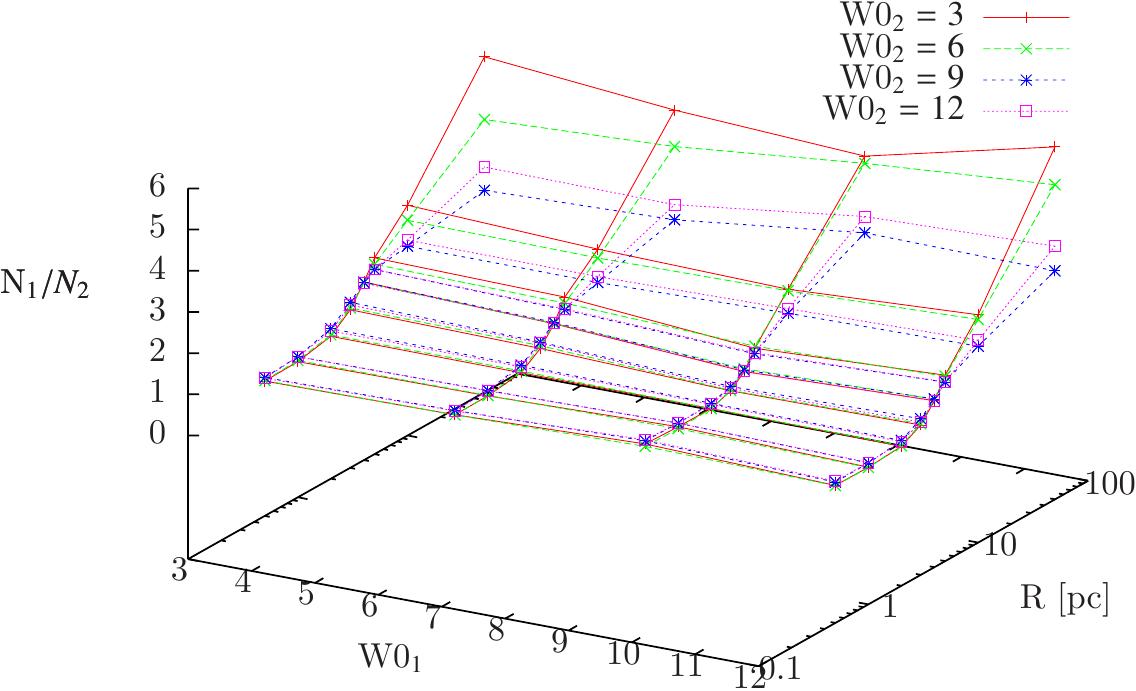}}
\caption
   {
    Similar to figure \ref{fig.fig_W1_constant} but for the cases in which $W0_2$ takes values 3 and 6,
    while $W0_1$ takes all values.
   }
\label{fig.fig_W2_constant}
\end{figure*}

After this first analysis, we now address the results for which the initial
King parameters are different. We depict in a three-dimensional figure the
final distribution of fractional occupation numbers as a function of the radius
and the King parameters in figures \ref{fig.fig_W1_constant} and
\ref{fig.fig_W2_constant}.  In the first one we keep $W0_1$ constant (set to 3,
6, 9 and 12) and we vary $W0_2$, and vice-versa in the
second figure. We can see that the King parameter only leaves a fingerprint for
the outer shells of the merged system in the first figure, and even more
remarkably on the second one, which also shows a more clear domination on the
number fraction of stars which initially belonged to Cluster 1. The two figures
are not symmetric in the distribution of $N_1/N_2$ along the radius because of
the initial difference in size, which is the dominant effect here. Since
Cluster 2 had initially half the size of Cluster 1 in all cases, only for low
values of W0 of Cluster 2 we can see a clear domination of $N_1$ over $N_2$.

\subsection{Equal-size clusters}
\label{sec.equal-size}

From the previous analysis we have seen that the size of the clusters very
likely plays a dominant role in the distribution of the different populations
as a function of the radius of the merged system. To shed light on this
dependence, we present in this section a second set of simulations which are
identical to the previous set presented before but for the radii of the two
clusters, which we fix to 3 pc in the two clusters.
We follow the structure of the first set of
simulations and use different W0 parameters ($W0_1$ and $W0_2$): 3, 6, 9 and 12
for the two clusters. This gives 16 possible combinations and this time we
perform two realisations per combination, which makes in total 32 simulations.
We set all other setup parameters (IMF, orbital parameters, age) identical to
the first set, so as to be able to understand the role of the size and King parameter more
clearly.

We run all simulations for T = 1200 $N-$body units, which corresponds to at
least one half-mass relaxation time in all cases.  In most of the simulations,
the two clusters merge before T=50 time units, in the sense that they share the
same density center, but the occupation fractions will still change
significantly after this time. We hence integrate the merged system further
until we reach at least $\sim 20$ times the merger time.

\begin{figure}
\resizebox{\hsize}{!}
          {\includegraphics[scale=1,clip]{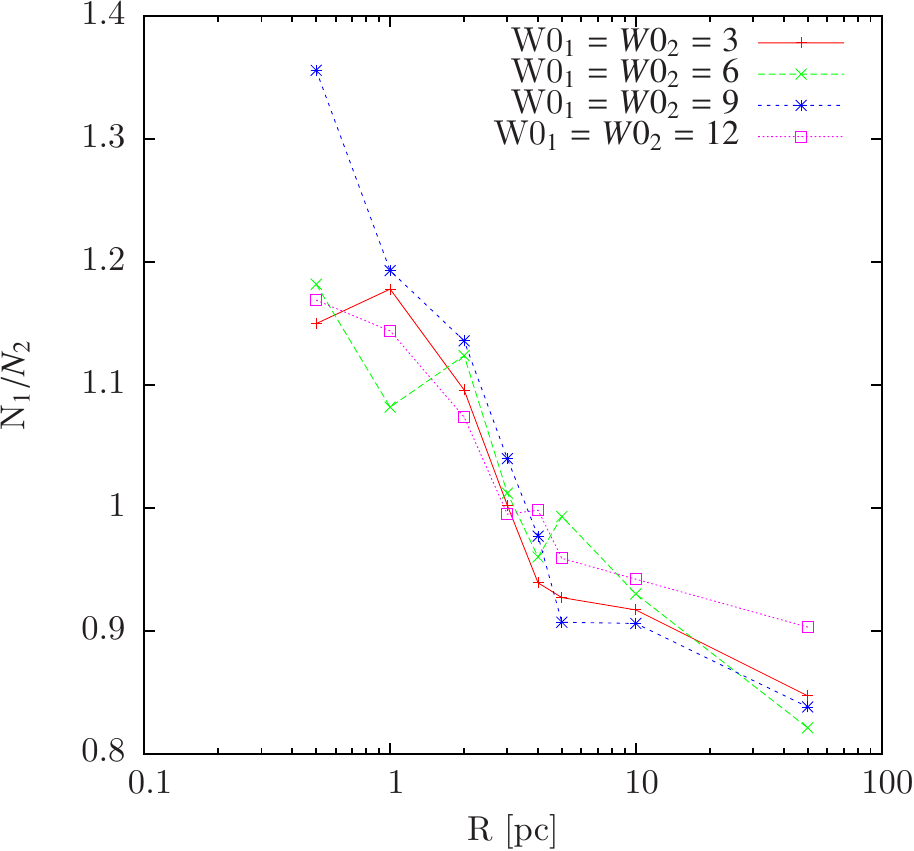}}
\caption
   {
    Fraction $N_1/N_2$ in different shells inside the final cluster at
    approximately $1.5\,T_{\rm rlx,\,h}$, for the cases where the two clusters have equal W0 King
parameters and initial sizes.
   }
\label{fig2.fig_equal_W}
\end{figure}

We first present the results for the simulations where the two
clusters have the same W0 parameters. The results are summarised in
Table \ref{tab2.table_equal_W} and Fig. \ref{fig2.fig_equal_W}.
We can see that in all cases $N_1/N_2 > 1$  at the centre, and the ratio progressively
decreases in the outer shells. This reveals the influence of the
different metallicity and age of the two clusters. Cluster 1 is
younger and has a lower metallicity (50 Myr old, Z = 0.01),
and thus it contains a larger number
of massive stars than Cluster 2 (100 Myr old, Z = 0.04).
Massive stars tend to concentrate at the centre of the final cluster,
because of mass segregation \citep[see e.g.][]{KhalEtAl07}. Thus, the centre is expected to
contain more stars of Cluster 1 than Cluster 2. Accordingly,
lower-mass stars populate mainly the outer parts of the system, thus
Cluster 2 dominates there as it contains more lower-mass stars.

We can see this more clearly by reading numbers in figure \ref{fig.IMFs}.
After stellar evolution, Cluster 1 has about 400
stars with $m>5 M_\odot$, while Cluster 2 has only about 110 stars. Also,
Cluster 1 has about 1,100 stars with $m>3 M_\odot$, while Cluster 2 only about
830. Since massive stars tend to concentrate at the centre, Cluster 1 dominates
there.
Finally, Cluster 1 has about 23,900 stars with $m<1 M_\odot$, while Cluster 2
has about 24,300 stars in the same mass-range. Those are low-mass stars that
are concentrated outside of the centre of the cluster, thus Cluster 2 dominates
in the outer parts of the final cluster.

\begin{table*}
\centering
\begin{tabular*}{14cm}{|p{2cm}|p{3cm}|p{3cm}|p{3cm}|p{3cm}|}
\cline{1-5}
\multicolumn{1}{|p{2cm}|}{\multirow{2}{*}{Shell  (pc)} }       &
  \multicolumn{4}{c|}{$N_1/N_2$}               \\ \cline{2-5}
 & $W0_1 = W0_2 = 3$ &$W0_1 = W0_2 = 6$ & $W0_1 = W0_2 = 9$& $W0_1 = W0_2 = 12$\\ \cline{1-5}
$0<r\leq 0.5$  &    1.150 $\pm$ 0.048
                        &     1.182 $\pm$ 0
                        &     1.356 $\pm$  0.014
                        &     1.169 $\pm$ 0.044 \\ \cline{1-5}

$0.5<r\leq 1$  &    1.178 $\pm$ 0.027
                        &    1.082 $\pm$ 0
                        &    1.193 $\pm$ 0.034
                        &    1.144 $\pm$ 0.0004 \\\cline{1-5}

$1<r\leq 2$     &    1.096 $\pm$ 0.007
                        &    1.124 $\pm$ 0
                        &    1.136 $\pm$ 0.008
                        &    1.074 $\pm$ 0.002      \\\cline{1-5}

$2<r\leq 3$     &    1.002 $\pm$ 0.009
                        &    1.012 $\pm$ 0
                        &    1.040 $\pm$ 0.015
                        &    0.995 $\pm$ 0.019        \\\cline{1-5}

$3<r\leq 4$     &   0.939 $\pm$ 0.011
                        &   0.960 $\pm$ 0
                        &   0.977 $\pm$ 0.013
                        &   0.998 $\pm$ 0.025          \\ \cline{1-5}

$4<r\leq 5$     &   0.927$\pm$ 0.044
                        &   0.993 $\pm$ 0
                        &   0.907 $\pm$ 0.005
                        &   0.959 $\pm$ 0.026 \\\cline{1-5}

$5<r\leq 10$   &   0.917 $\pm$ 0.004
                        &   0.930 $\pm$ 0
                        &   0.906 $\pm$ 0.020
                        &   0.942 $\pm$ 0.019   \\\cline{1-5}

$10<r\leq 50$  &  0.847 $\pm$ 0.01
                         &  0.821 $\pm$ 0
                         &  0.838 $\pm$ 0.024
                         &  0.903 $\pm$ 0.024 \\\cline{1-5}
\end{tabular*}
\caption{
Summary of the results for for simulations with $W0_1 = W0_2$ for clusters that initially have the same size.
}
\label{tab2.table_equal_W}
\end{table*}

\begin{figure}
\resizebox{\hsize}{!}
          {\includegraphics[scale=1,clip]{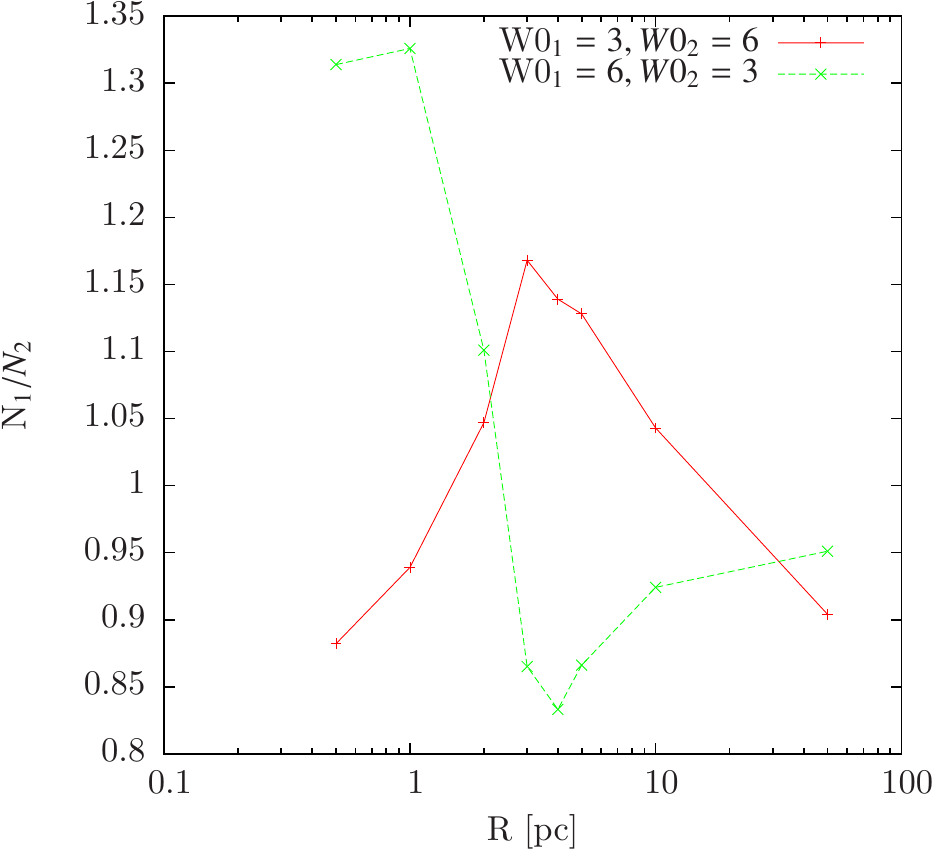}}
\caption
   {
    Same as in figure \ref{fig2.fig_equal_W} for the cases where the two clusters have $W0_1 =
    3$, $W0_2=6$ and $W0_1 = 6$, $W0_2=3$.
   }
\label{fig2.figure_W_3_6}
\end{figure}

\begin{figure}
\resizebox{\hsize}{!}
          {\includegraphics[scale=1,clip]{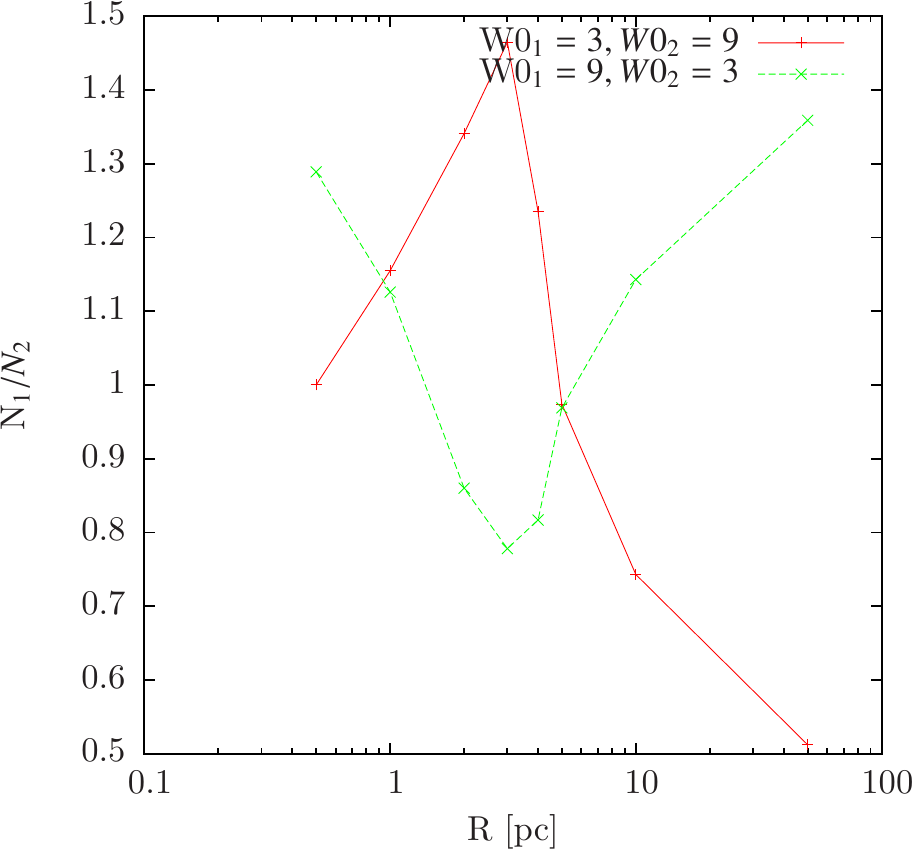}}
\caption
   {
    Same as in figure \ref{fig2.fig_equal_W} for the cases where the two clusters have $W0_1 =
    3$, $W0_2=9$ and $W0_1 = 9$, $W0_2=3$.
   }
\label{fig2.figure_W_3_9}
\end{figure}

\begin{figure}
\resizebox{\hsize}{!}
          {\includegraphics[scale=1,clip]{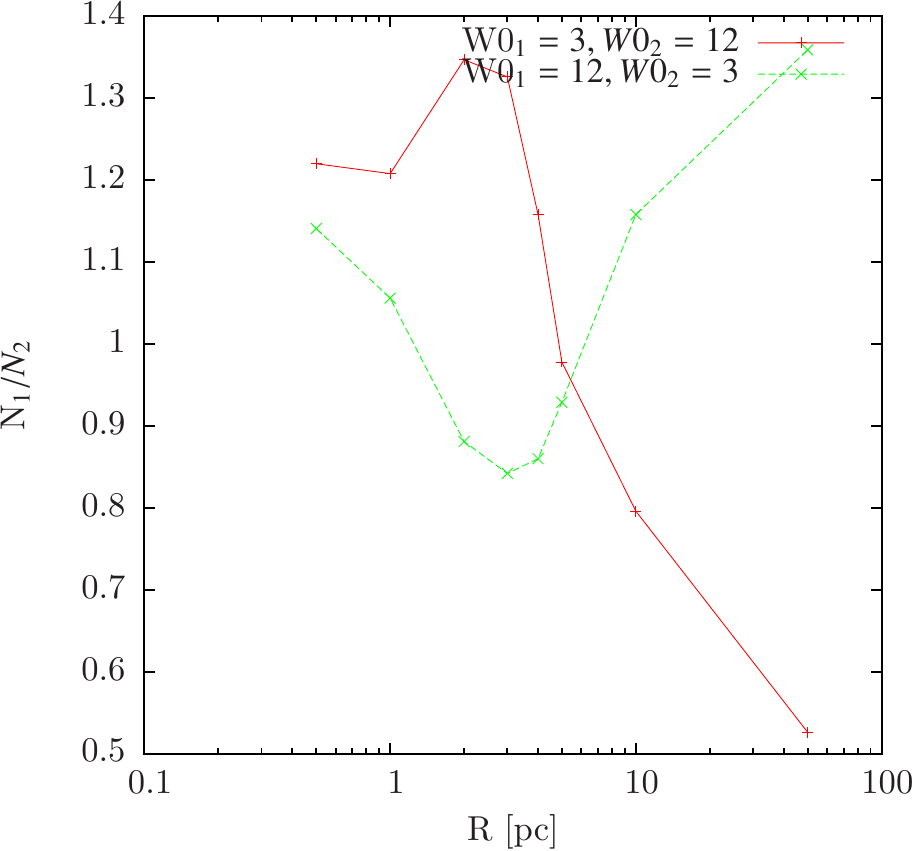}}
\caption
   {
    Same as in figure \ref{fig2.fig_equal_W} for  $W0_1 =
    3$, $W0_2=12$ and $W0_1 = 12$, $W0_2=3$.
   }
\label{fig2.figure_W_3_12}
\end{figure}

\begin{figure}
\resizebox{\hsize}{!}
          {\includegraphics[scale=1,clip]{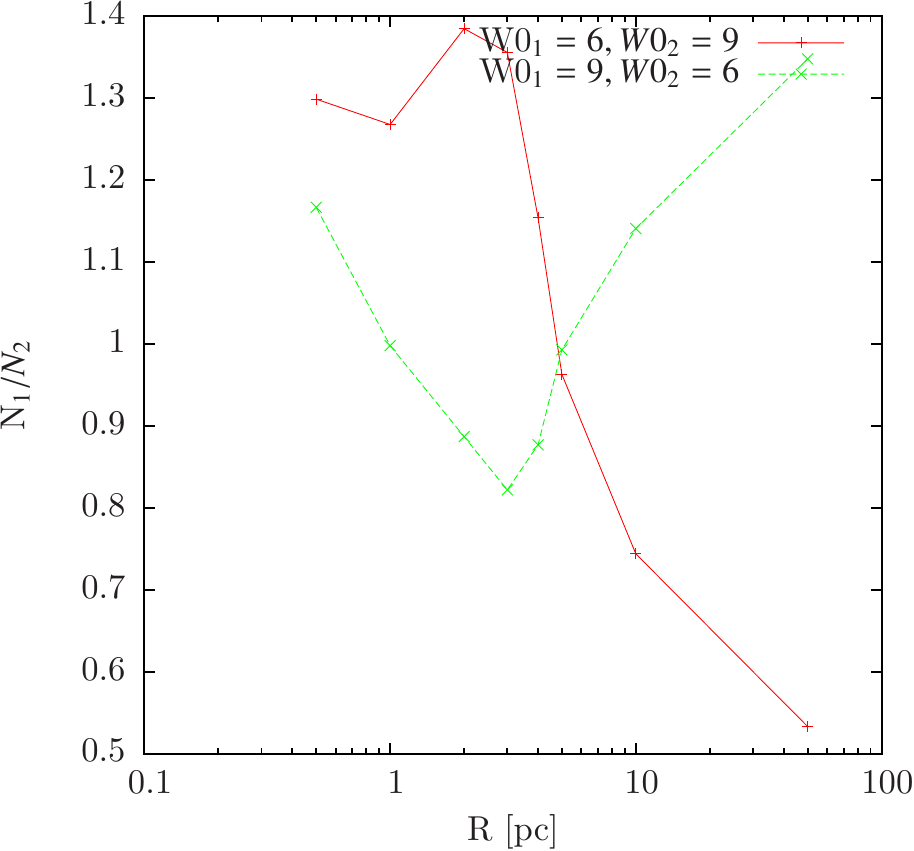}}
\caption
   {
    Same as in figure \ref{fig2.fig_equal_W} but for $W0_1 =
    6$, $W0_2=9$ and $W0_1 = 9$, $W0_2=6$.
   }
\label{fig2.figure_W_6_9}
\end{figure}

\begin{figure}
\resizebox{\hsize}{!}
          {\includegraphics[scale=1,clip]{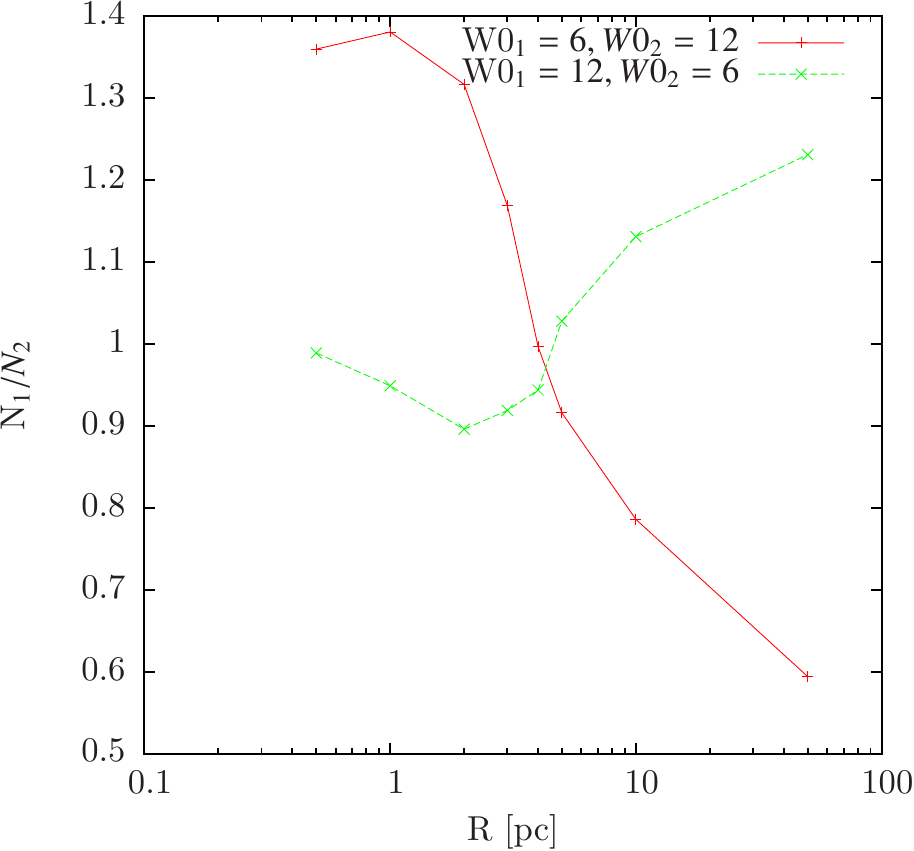}}
\caption
   {
    Same as in figure \ref{fig2.fig_equal_W} but here we have $W0_1 =
    6$, $W0_2=12$ and $W0_1 = 12$, $W0_2=6$.
   }
\label{fig2.figure_W_6_12}
\end{figure}

\begin{figure}
\resizebox{\hsize}{!}
          {\includegraphics[scale=1,clip]{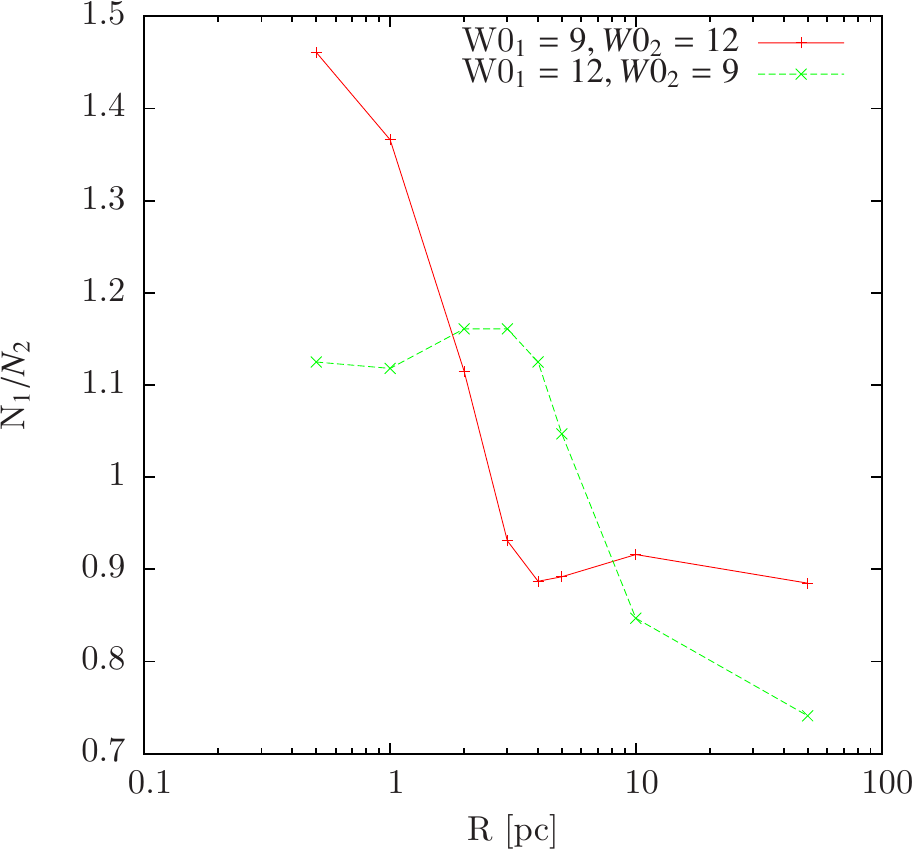}}
\caption
   {
    Same as in figure \ref{fig2.fig_equal_W} for $W0_1 =
    9$, $W0_2=12$ and $W0_1 = 12$, $W0_2=9$.
   }
\label{fig2.figure_W_9_12}
\end{figure}

We now address the simulations in which $W0_1 \neq W0_2$. The aim here is to
investigate the influence of the King parameter in the final radial
distribution of the two populations. In figures
\ref{fig2.figure_W_3_6} -- \ref{fig2.figure_W_9_12} we present the results of
simulations for which we have taken $W0_1 = X, W0_2=Y$ and $W0_1 = Y, W0_2=X$,
with $X \neq Y$. The results in most of the cases show that the two lines (line
for $W0_1<W0_2$ (red) and line for $W0_1>W0_2$ (green)) are almost mirror
copies of each other.

In all cases except where $X=9$ and $Y=12$, the $W0_1<W0_2$ line shows its
maximum at about the same distance from the centre at which the $W0_1>W0_2$
shows its minimum. The distance where the lines show their peaks ranges from 1
pc to 3 pc, which is close to the half-mass radius of the cluster.  The biggest
difference between the two peaks is found in the case $X=3$, $Y=9$.  In the
exceptional case with $X=9$ and $Y=12$, as we discussed before, the two lines
are again almost mirror copies of each other, but this time, contrary to all
other cases, the very dilute core results in a $W0_1<W0_2$ line that shows a
minimum while the $W0_1>W0_2$ is a maximum.

In all cases  except those in which $X=3$, $Y=9$ and $X=9$, $Y=12$, the outer parts of the
cluster are dominated by Cluster 2, for $W0_1<W0_2$ (the red lines have
$N1/N2<1$ in the outer shells), while Cluster 1 dominates the outer parts for
$W0_1>W0_2$ (the green lines have $N1/N2>1$ in the outer shells). In the inner
parts of the cluster the situation is more complicated, since the influence of
the King parameters is mixed with the influence of the stellar evolution and
the fact that the massive stars of Cluster 1 are more numerous than those of
Cluster 2.

\section{A study of the rotation}
\label{sec.rot}

\begin{figure*}
\resizebox{0.8\hsize}{!}
          {\includegraphics[scale=1,clip]{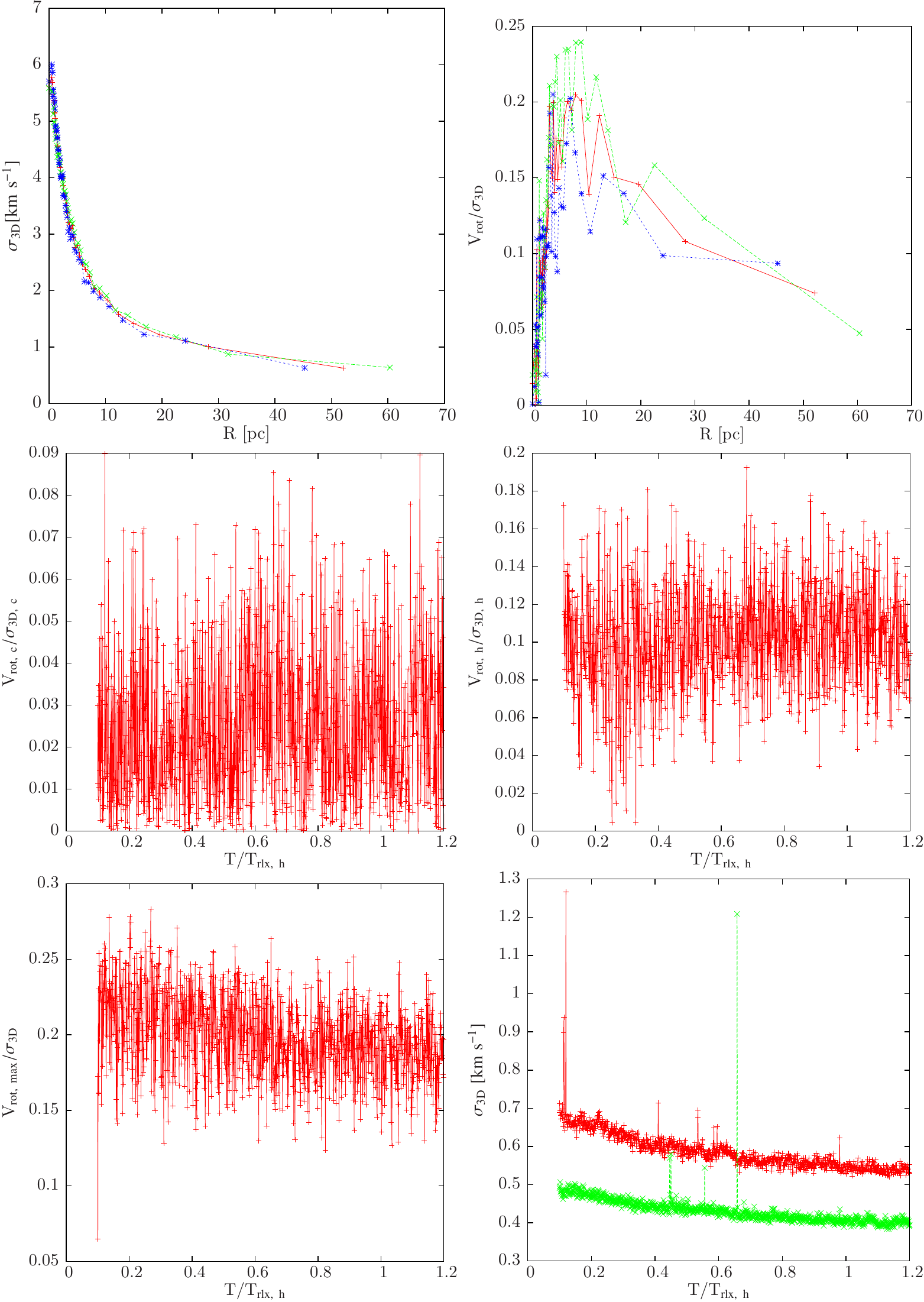}}
\caption
   {
{From the left to the right and from the top to the bottom and for the
simulation in which the clusters had initially the same size and King
parameters 9 and 6, we depict the following quantities: The three-dimensional
velocity dispersion $\sigma_{\rm 3D}$ of the merged cluster as a function of
the radius $R$ at the end of the simulation, which happens at $1.2\,T_{\rm rlx}$, the rotational velocity $V_{\rm rot}$ of the merged
cluster normalised to $\sigma_{\rm 3D}$ as a function of $R$. The evolution of
$V_{\rm rot}$ at the core radius, $V_{\rm rot,\,c}$ normalised to the core
three-dimensional velocity dispersion $\sigma_{\rm 3D,\,c}$, the evolution of
the half-mass rotational velocity $V_{\rm rot,\, h}$ normalised to the
half-mass three-dimensional velocity dispersion $\sigma_{\rm 3D,\,h}$, the
evolution of the maximum rotational velocity $V_{\rm rot,\,max}$ normalised to
$\sigma_{\rm 3D}$ and the evolution of $\sigma_{\rm 3D}$. The solid, red lines
correspond to the total merged cluster, while we show stars which originally belonged
to Cluster 1 in dashed, blue lines with star symbols and in dashed, greed lines with
crosses to Cluster 2.
}
   }
\label{fig.Rotation_9_6_2}
\end{figure*}

{In this subsection we present a brief analysis of the rotation of the merged clusters for
the case in which they initially had the same size. We study the rotation as a function of
the radius by calculating different quantities. In figure~\ref{fig.Rotation_9_6_2} we show a
set of plots for a representative case, in particular the second iteration of the simulation in which initially $W0_1=9$ and
$W0_2=6$. The fact that the clusters have merged is reflected in the first panel, since the three 3-D velocities dispersions
are distributed very similarly over the radius of the final cluster.
The maximum
value is reached at about $\sim 10$ pc from the density centre, resulting in a significant
rotation even after $1.2\,T_{\rm rlx}$.
The next two panels show the rotational velocity evolution normalised to the
velocity dispersion at two relevant radii, the core radius and the half-mass
radius. The second panel of the middle row shows us that, although with
fluctuations, the high values reached at the inner parts of the cluster are
kept during the evolution of the simulations, to achieve values between 0.07
and 0.12 after $1.2\,T_{\rm rlx}$. Although it would be important to study the
evolution of rotation over longer time spans, we cannot afford it with our
code.
The last two panels show the maximum rotational velocity as a function of time, which
shows a clear slow decay towards lower values and the evolution of the 3D velocity
dispersion in the cluster.
}

\begin{table*}
    \centering
    \begin{tabular}{|c|c|c c c c|c|}
        \hline
         \multicolumn{2}{|c|}{\multirow{2}{*}{$V_{\rm rot}/{\sigma}$}}&\multicolumn{4}{c|}{W0$_1$}\\
        \cline{3-6}
        \multicolumn{2}{|c|}{}&              3             &        6                 &           9          &      12                \\
        \hline
        \multirow{12}{*}{W0$_2$}
             & \multirow{3}{*}{3} & $0.090\pm 0.011$ &  $0.042\pm 0.012$    & $0.037\pm 0.029$ & $0.043\pm 0.013$\\
             &                    & $0.149\pm 0.036$ &  $0.136\pm 0.016$    & $0.120\pm 0.023$ & $0.117\pm 0.005$\\
             &                    & $0.186\pm 0.017$ &  $0.178\pm 0.026$    & $0.164\pm 0.038$ & $0.153\pm 0.001$\\
             \cline{3-6}
             & \multirow{3}{*}{6} & --                     &  $ 0.046 \pm  0.003$ & $0.009 \pm 0.004$  & $0.026\pm 0.001$ \\
             &                    & --                     &  $ 0.138 \pm  0.024$ & $0.060 \pm 0.012$  & $0.154\pm 0.040$ \\
             &                    & --                     &  $ 0.224 \pm  0.003$ & $0.151 \pm 0.020$  & $0.195\pm 0.001$  \\
             \cline{3-6}
             & \multirow{3}{*}{9} & --                     &  --                         &$ 0.012\pm 0.008$  & $0.039 \pm 0.005$ \\
             &                    & --                     &  --                         &$ 0.100\pm 0.005$  & $0.108 \pm 0.028$ \\
             &                    & --                     &  --                         &$ 0.227\pm 0.007$  & $0.189 \pm 0.019$ \\
             \cline{3-6}
             & \multirow{3}{*}{12}& --                     &  --                         &  --                     & $0.029 \pm 0.008$   \\
             &                    & --                     &  --                         &  --                     & $0.062 \pm 0.003$   \\
             &                    & --                     &  --                         &  --                     & $0.182 \pm 0.061$   \\
        \hline
    \end{tabular}
    \caption{{$V_{\rm rot}/{\sigma}$ at different radii for the different King parameters considered, in the case of same sizes.
             In each cell we display the value
             at the core radius, $R_{\rm c}$, the half-mass radius, $R_{\rm h}$ and the radius in which the merged cluster delivers the
             maximum value of $V_{\rm rot}$, $R_{\rm V_{max}}$, from the top to the bottom respectively, along with the deviation that
             we obtained from the corresponding number of simulations for every combination of King parameters.}}
    \label{t.VrotSigma}
\end{table*}

\section{The particular case of $\omega$ Cen}
\label{sec.omecen}

\begin{figure*}
\resizebox{\hsize}{!}
          {\includegraphics[scale=1,clip]{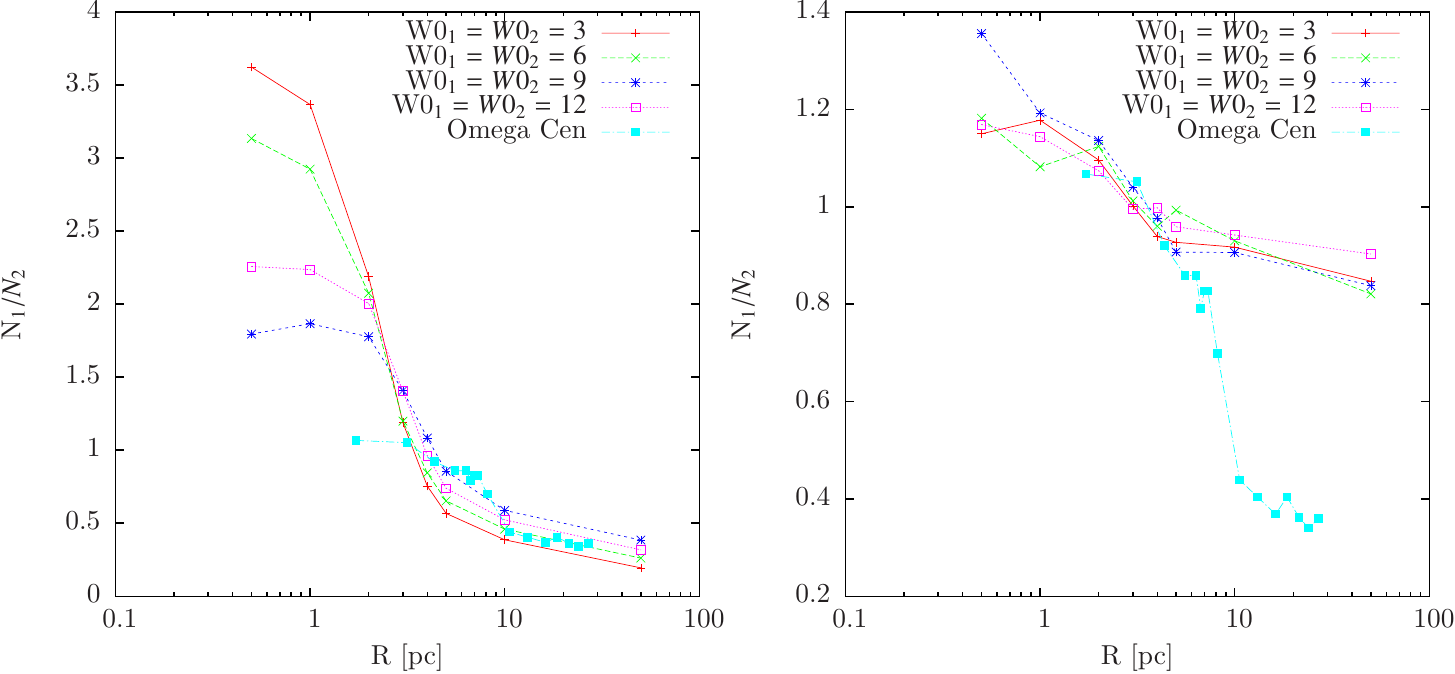}}
\caption
   {
{Fractional occupation numbers as a function of the radius as observed in $\omega$ Cen (cyan, solid
curve with squares), compared to our models for clusters that initially had the same size, left
panel, and in the right panel clusters that formed out of the merger of two clusters with different radii, as explained
in section~\ref{sec.results2}.}
   }
\label{fig.ComparOmegaCen}
\end{figure*}

{In our simulations we are limited in our study of the vast parameter space
by our code. While direct-summation techniques are very robust, they scale
typically as $\propto N^2$. Moreover, the approach we use to evolve the
clusters and their metallicities is rather simplistic compared to what happens
in real clusters. Nevertheless, in figure~\ref{fig.ComparOmegaCen} we present a
comparison of our scenario with observations of $\omega$ Cen, in particular with the data
presented in the work of \cite{BelliniEtAl09}.
In the left panel of the figure we can see that the observed data of $\omega$ Cen
follows the trend of the shape of the rest of curves at smaller radii. While we
are not limited in our resolution at smaller radii, observations are but, on the
other hand, we are comparing a very massive cluster, actually an ultra-compact
dwarf galaxy with a model of two clusters of 30,000 stars each. At larger radii,
the observed data of $\omega$ Cen suits best the model in which the two clusters
had a King parameter of 6. On the right panel we observe the same discrepancy that
we already had in our figures~\ref{fig.fig_equal_W} and \ref{fig2.fig_equal_W} at
larger radii. Therefore it seems that $\omega$ Cen was probably formed out of the merger
of clusters that initially had a similar size, mass and King parameter.
Nonetheless, the star resolution is well below the number of stars that we expect in $\omega$ Cen,
and the simulations might depend on the masses and sizes of the simulated clusters, so that this result should
be taken carefully but, in any case, is an encouraging motivation for further probing this scenario.
}

\section{Summary}
\label{sec.summary}

Motivated by the seemingly unavoidable future merger of clusters with different
ages --and hence with potentially different metallicities-- in the Antenn{\ae}
galaxy, in this paper we have presented direct-summation $N$-body simulations
of such a merger process. Such mergers were addressed in the past by different
authors, but ours is the first study to explicitly tackle the possibility of
multiple metallicities being present in the merging clusters.

Interacting galaxies such as the Antenn{\ae} are natural loci for multi-metallic
clusters to collide. In the Antenn{\ae}, CCs have been observed with the HST
and they are the natural birth-place of UCDs, as explained previously.
While dynamics does not affect the shape of the resulting CMD of the merged
system, it impinges the number of stars in different radial regions of the
resulting merger: the occupation fraction will vary with radius, because of
the dynamics.

In the first part of this article we have run a sample of simulations in which
we vary all relevant parameters for the final distribution of stars in the
merged cluster. We conclude that the initial concentration of the clusters,
metallicities, initial ages and sizes play an important role in the final
distribution.  The differences in the features of the original clusters may
well be observed in the CMD of the final cluster, which consists of multiple
lines, which might appear merged together in the lower part of the diagram, but
could be clearly distinguishable in its upper part, even if the distance to the
cluster is up to 5 Mpc. We note that our simulations show that clusters that
are created by mergers of smaller clusters exhibit phases with ellipticity
above the observational average. This indicates that future observations of
young clusters hosting multiple stellar populations should focus on oblate
cluster with high rotation.

In the second part of the paper we perform an exhaustive analysis of the role
of the King parameter and the size of the clusters in the final distribution of
stars. We adapt case ${\cal A}$ as our fiducial scenario and fix the number of
stars, the metallicity and age.

We start by fixing the size of the clusters, so that one is twice as large
as the other one and run 128 simulations {(of 8 realisations of each
combination of W0 King parameters, to vary the random seed)}. We find that the
dominant parameter on the final distribution of stars is the initial difference
of sizes of the clusters.  Almost totally independent on the rest of parameters
the occupation fractional number is $N_1/N_2 < 1$ closer to the centre of the
cluster and $>1$ outside in all simulations.

We then investigate the role of the metallicity and age by running a set of 32 simulations
in which the two clusters have equal sizes. In most of the cases, and
independently of W0, the core of the final cluster is dominated by stars from
Cluster 1. However, the outer parts of the systems are mostly affected by the
initial difference in the choice for W0.

Although we cannot add abundance ratios for many different chemical species to
our models, our analysis provides guidance into some possible observational
signatures of merger events, including the possible presence of increased
rotation and high ellipticity, associated with the existence of multiple
metallicities in individual, present-day GCs.

{In particular, it is remarkable to see that the set of models in which the
clusters had initially the same size leads to a distribution of occupational fractions
rather close to what is {\em observed} in $\omega$ Cen. Therefore, in this scenario in
which different populations are a fingerprint of cluster mergers, $\omega$ Cen was formed
out of collisions among clusters with relatively similar sizes, which is reasonable, since
$\omega$ Cen has been proposed to have formed in a cluster complex dynamically, and the
clusters that lead to the formation of a runanay seed ultra-compact dwarf galaxy have a
similar mass and size, due to mass segregation, as in the work of \cite{AmaroSeoaneEtAUCDs}}

\section*{Acknowledgments}

It is a pleasure to thank Marc Freitag, Andrea Bellini and Pavel Kroupa for
comments on the manuscript. We are indebted to Jarrod R.  Hurley for his
assistance with the stellar evolution {\tt sse} package, as well as to Peter
Teuben with {\tt Nemo}. We acknowledge financial support for research visits in
China by The Silk Road Project (2009S1-5) of Chinese Academy of Sciences,
National Astronomical Observatories of China.  We are thankful to Christoph
Olczak for his support with the GPU nodes at the NAOC/CAS and the Titan cluster
in Heidelberg, as well to Rainer Spurzem for granting access to them.  This
work has been supported by the Transregio 7 ``Gravitational Wave Astronomy''
financed by the Deutsche Forschungsgemeinschaft DFG (German Research
Foundation).  PAS was supported in part by the National Science Foundation
under Grant No. 1066293 and he thanks the hospitality of the Aspen Center for
Physics, as well as Jorge Cuadra for his invitation to the Universidad
Cat{\'o}lica de Chile. S.K. was supported by the Deutsches
Zentrum f{\"u}r Luft- und Raumfahrt and by
the FONDECYT Postdoctoral Fellowship number 3130570. M.C. is
supported by Proyecto Fondecyt Regular \#1110326; BASAL Center for Astrophysics
and Associated Technologies (PFB-06); FONDAP Center for Astrophysics
(15010003); the Chilean Ministry for the Economy, Development, and Tourism's
Programa Iniciativa Cient\'{i}fica Milenio through grant P07-021-F, awarded to
The Milky Way Millennium Nucleus; and Proyecto Anillo ACT-86. We thank Emma
Robinson for comments on the manuscript.  We acknowledge Ant{\'o}n Ricard Amaro
Pendl for suggesting us to plot in B\&W the clusters in the movies.

\label{lastpage}

\end{document}